\documentclass[prd,nofootinbib,showpacs]{revtex4}
\usepackage{graphics,color}
\usepackage{bm}
\usepackage{graphicx}
\usepackage{amsmath}
\usepackage{amssymb}
\usepackage{enumitem}
\usepackage{tensor}	
\usepackage[fleqn]{mathtools}
\usepackage{epstopdf}
\usepackage{hyperref}
\allowdisplaybreaks	

\def\be{\begin{equation}}
\def\ee{\end{equation}}
\def\ba{\begin{eqnarray}}
\def\ea{\end{eqnarray}}
\def\l{\left}
\def\r{\right}
\def\f{\frac}

\def\nn{\nonumber}

\def\hub{{\mathcal H}}

\newcommand{\SDer}{\bar{\nabla}}

\def\ie{{\frenchspacing\it i.e. }}

\begin{document}

\title{EFTCAMB/EFTCosmoMC: Numerical Notes v3.0}
\author{Bin Hu$^{1}$, Marco Raveri$^{2,3}$, Noemi Frusciante$^{4}$, Alessandra Silvestri$^{3}$}
\smallskip
\affiliation{$^{1}$ Department of Astronomy, Beijing Normal University, Beijing, 100875, China\\
\smallskip 
$^{2}$ Kavli Institute for Cosmological Physics, Department of Astronomy \& Astrophysics,
Enrico Fermi Institute, The University of Chicago, Chicago, IL 60637, USA\\
\smallskip
$^{3}$  Institute Lorentz, Leiden University, PO Box 9506, Leiden 2300 RA, The Netherlands \\
\smallskip
$^{4}$ Instituto de Astrof$\acute{\text{i}}$sica e Ci$\hat{\text{e}}$ncias do Espaco, 
Faculdade de Ci$\hat{\text{e}}$ncias da Universidade de Lisboa, Lisbon, Portugal}

\begin{abstract}
EFTCAMB/EFTCosmoMC are publicly available patches to the CAMB/CosmoMC codes implementing the effective field theory approach to single scalar field dark energy and modified gravity models. With the present numerical notes we provide a guide to the technical details of the code. Moreover we reproduce, as they appear in the code, the complete set of the modified equations and the expressions for all the other relevant quantities used to construct these patches. We submit these notes to the arXiv to grant full and permanent access to this material which provides very useful guidance to the numerical implementation of the EFT framework. We will update this set of notes when relevant modifications to the EFTCAMB/EFTCosmoMC codes will be released. 

The present version is based on the version of EFTCAMB/EFTCosmoMC$\_$Sep17. 
\end{abstract}

{ 
\maketitle

\tableofcontents

\section{Introduction}\label{Sec:Intro}
In the quest to address one of the most pressing problems of modern cosmology, \ie cosmic acceleration, an effective field theory approach has been recently proposed~\cite{Gubitosi:2012hu,Bloomfield:2012ff}. The virtue of this approach relies in the  model-independent description of this phenomenon as well as in the possibility to cast into the EFT language most of the single field DE/MG gravity models of cosmological interest~\cite{Gubitosi:2012hu,Bloomfield:2012ff,Gleyzes:2013ooa,Bloomfield:2013efa}.  The EFT action is written in unitary gauge and Jordan frame and it contains all the operators invariant under time-dependent spatial diffeomorphisms, ordered in power of perturbations and derivatives. These operators enter in the action with a time dependent function in front of them,  to which we will refer to as EFT functions. The DE/MG models encoded in this formalism have one extra scalar d.o.f. and a well defined Jordan frame; in unitary gauge the scalar field is hidden in the metric. In order to study the 
dynamics of scalar perturbations, it is better to make its dynamics manifest via the St$\ddot{\text{u}}$ckelberg technique, \ie restoring  the time diffeomorphism invariance through an infinitesimal time coordinate transformation. Then a  new scalar field $\pi$ appears in the action, the so called St$\ddot{\text{u}}$ckelberg field. \\
The EFT action in conformal time reads
\begin{align}\label{full_action_Stuck}
S = \int d^4x& \sqrt{-g} \left \{ \frac{m_0^2}{2} \l[1+\Omega(\tau+\pi)\r]R+ \Lambda(\tau+\pi) - c(\tau+\pi)a^2\left[ \delta g^{00}-2\frac{\dot{\pi}}{a^2} + 2\hub\pi\left(\delta g^{00}-\frac{1}{a^2}-2\frac{\dot{\pi}}{a^2}\right) +2\dot{\pi}\delta g^{00} \right.\right. \nonumber \\
 &\left.\left.+2g^{0i}\partial_i\pi-\frac{\dot{\pi}^2}{a^2}+ g^{ij}\partial_i \pi \partial_j\pi -\l(2\hub^2+\dot{\hub}\r)\frac{\pi^2}{a^2} + ... \right] \right. \nonumber \\
 &\left. + \frac{M_2^4 (\tau + \pi)}{2}a^4 \left(\delta g^{00} - 2 \frac{\dot{\pi}}{a^2}-2\frac{\hub\pi}{a^2}+... \right)^2 \right. \nonumber\\
& \left.- \frac{\bar{M}_1^3 (\tau + \pi)}{2}a^2 \left(\delta g^{00}- 2 \frac{\dot{\pi}}{a^2}-2\frac{\hub\pi}{a^2}+... \right) \left(\delta 
\tensor{K}{^\mu_\mu} + 3 \frac{\dot{\hub}}{a} \pi + \frac{\SDer^2 \pi}{a^2}+ ...\right) \right. \nonumber \\
&\left. - \frac{\bar{M}_2^2 (\tau + \pi)}{2} \left(\delta \tensor{K}{^\mu_\mu} + 3 \frac{\dot{\hub}}{a} \pi + \frac{\bar{\nabla}^2 \pi}{a^2} + ... \right)^2 \right.\nonumber  \\
& \left. - \frac{\bar{M}_3^2 (\tau + \pi)}{2}
  \left(\delta \tensor{K}{^i_j} + \frac{\dot{\hub}}{a} \pi \delta\indices{^i_j}
  + \frac{1}{a^2} \bar{\nabla}^i \bar{\nabla}_j \pi +... \right)
  \left(\delta \tensor{K}{^j_i} + \frac{\dot{\hub}}{a} \pi \delta\indices{^j_i}
  + \frac{1}{a^2} \SDer^j \SDer_i \pi + ... \right) \nonumber \right. \\
&  \left. + \frac{\hat{M}^2 (\tau + \pi)}{2} a^2 \left(\delta g^{00} - 2 \frac{\dot{\pi}}{a^2} - 2\frac{\hub}{a^2}\pi + ... \right)\, \left(\delta R^{(3)} +4\frac{\hub}{a} \bar{\nabla}^2\pi + ...\right)\right. \nonumber \\
&\left. + m_2^2(\tau+\pi)\left(g^{\mu\nu}+n^{\mu} n^{\nu}\right)\partial_{\mu}\left(a^2g^{00}-2\dot{\pi} -2\hub\pi + ...\right)\partial_{\nu}\left(a^2g^{00}-2\dot{\pi} - 2\hub\pi + ...\right) + ...\right\}  + S_{m} [g_{\mu \nu},\chi_i],
\end{align}
where $m_0^2$ is the Planck mass, overdots represent derivatives with respect to conformal time and $\bar{\nabla}$ indicates three dimensional spatial derivatives. $\{\Omega$,$\Lambda$,$c\}$ are the only three EFT functions describing the background dynamics, hence the name background functions. While the dynamics of linear scalar perturbations is described by the second order EFT functions, $\{M_2,\bar{M}_1,\bar{M}_2,\bar{M}_3,\hat{M},m_2\}$, in combination with the background ones. We parametrize the conformal coupling to gravity via the function $1+\Omega$ instead of $\Omega$~\cite{Gubitosi:2012hu,Bloomfield:2012ff} for reasons of numerical accuracy. Finally, $S_m$ is the action for all matter fields, $\chi_i$. The EFT approach relies on the assumption  of the validity of  the weak equivalence principle which ensures the existence of a metric universally coupled to matter fields and therefore of a well defined Jordan frame.

In~\cite{Hu:2013twa,Raveri:2014cka}, we introduced EFTCAMB which is a patch of  the publicly available  Einstein-Boltzmann solver, CAMB~\cite{CAMB,Lewis:1999bs}. The code implements the EFT approach, allowing to study the linear cosmological perturbations  in a model-independent framework via the \textit{pure} EFT procedure, although it ensures to investigate the dynamics of linear perturbations of specific single scalar field DE/MG models via the \textit{mapping} EFT procedure, once the matching is worked out. EFTCAMB evolves the full perturbation equations on all linear scales without relying on any quasi static approximation. Moreover it checks the stability conditions of perturbations  in the dark sector in order to ensure that the underlying gravitational theory is acceptable.
Finally, it enables to specify the expansion history by choosing a DE equation of state among several common parametrizations, allowing phantom-divide crossings.
To interface EFTCAMB with cosmological observations we equipped it with a modified version of 
CosmoMC~\cite{Lewis:2002ah}, what we dub EFTCosmoMC~\cite{Raveri:2014cka}. EFTCosmoMC  allows to practically perform tests of gravity and get constraints on the parameter space using cosmological data sets. The stability conditions implemented in EFTCAMB translate into EFTCosmoMC  as {\it viability priors} to impose  on parameters describing the dark sector. The first release of the code includes data such as \textit{Planck}, WP, BAO and \textit{Planck} lensing. The present version is fully compatible with all data sets available in CosmoMC. The EFTCAMB/EFTCosmoMC package is now publicly available for download at~\url{http://www.eftcamb.org}.

After EFTCAMB some other general purpose Einstein-Boltzmann codes modeling scalar-tensor theories have been developed, such as hi-class~\cite{Zumalacarregui:2016pph} and COOP~\cite{Huang:2015srv}.  In a recent paper~\cite{Bellini:2017avd} it has been shown that EFTCAMB (Sep17) and hi-class  agree to a high level of accuracy, while their agreement with COOP is sufficiently high only on large scales. It has been also shown that the implementation of the low-energy Ho\v rava gravity model in EFTCAMB gives the same results of the LVDM CLASS code~\cite{Blas:2012vn}.

Throughout this Numerical Notes we will always use the following conventions:
\begin{itemize}
\item The overdot represents derivation with respect to conformal time $\tau$ while the prime represents derivation with respect to the scale factor $a$, unless otherwise specified.
\item In what follows we define a new dimensionless St$\ddot{\text{u}}$ckelberg field: $\widetilde{\pi}$,  \ie the $\pi$-field in the action~(\ref{full_action_Stuck}) multiplied by $H_0$ and divided by $a$. For the rest of the notes we will suppress the tilde to simplify the equations so $\widetilde{\pi}$ is written as $\pi$, if there is no confusion.
\item We redefine all the second order EFT functions to make them dimensionless and to facilitate their inclusion in the code: 

\begin{align}
&\gamma_1=\frac{M^4_2}{m_0^2H_0^2}, \,\,\,\,\,\gamma_2=\frac{\bar{M}^3_1}{m_0^2H_0}, \,\,\,\,\,\gamma_3=\frac{\bar{M}^2_2}{m_0^2}, \nonumber \\
&\gamma_4=\frac{\bar{M}^2_3}{m_0^2}, \,\,\,\,\,\gamma_5=\frac{\hat{M}^2}{m_0^2},\,\,\,\,\,\gamma_6=\frac{m^2_2}{m_0^2}.
\end{align}
Let us notice that after v2.0 we have slightly changed the definition of the second order EFT functions w.r.t. the convention used in v1.0 and v1.1. These new definitions do not change the general structure and physics in the code, but they allow for a more direct and cleaner implementation of Horndeski models~\cite{Horndeski:1974wa,Deffayet:2009mn}. For the sake of clarity let us list the explicit correspondence between this new convention and that one used in the previous releases (v1.0 and v1.1), in terms of the $\alpha$'s:
\be
\gamma_1=\alpha_1^4, \,\,\,\,\,\gamma_2=\alpha_2^3, \,\,\,\,\,\gamma_3=\alpha_3^2,\,\,\,\,\,
\gamma_4=\alpha_4^2, \,\,\,\,\,\gamma_5=\alpha_5^2,\,\,\,\,\,\gamma_6=\alpha_6^2.
\ee

\item We define all the EFT functions $\Omega$, c, $\Lambda$ and the $\gamma$-functions as function of the scale factor $a$ .
\end{itemize}
%
\section{The structure of the code}\label{Sec:Code structure}
The structure of the EFTCAMB code is illustrated in the flowchart of Figure~\ref{Fig:StructureFlowChart}. There is a number associated to each model selection flag; such number is reported in Figure~\ref{Fig:StructureFlowChart} and it controls the behaviour of the code.
The main code flag is \texttt{EFTflag}, which is the starting point after which all the other sub-flags, can be chosen according to the user interests.
\begin{itemize}
\item The number $\texttt{EFTflag}=0$ corresponds to the standard CAMB code. Every EFT modification to the code is automatically excluded by this choice.
\item  The number $\texttt{EFTflag}=1$ corresponds to \textit{pure} EFT models. 
The $\texttt{PureEFTmodel}$ flag controls which basis for pure EFT to use.
For the choice $\texttt{PureEFTmodel}=1$ the user needs to select a model for the background expansion history via the \texttt{EFTwDE} flag. Various common parametrizations for the DE equation of state are natively included in the code.
The implementation details of the dark energy equations of state can be found in Section~\ref{SubSec:DeEOS}.
Finally, to fully specify the {\it pure} EFT model, one has to fix the EFT functions behaviour as functions of the scale factor $a$. The corresponding flags for the model selection are: \texttt{PureEFTmodelOmega} for the model selection of the EFT function $\Omega(a)$ and \texttt{PureEFTmodelGamma}$\text{i}$, with $\text{i}=1,..,6$ for the $\gamma_i(a)$ EFT functions. Some built-in models are already present in the code and can be selected with the corresponding number, see Flowchart~\ref{Fig:StructureFlowChart}. The details about these models can be found in Section~\ref{Sec:PureEFTmodels}.
There is also the possibility to use the flag \texttt{PureEFTHorndeski} to restrict \textit{pure} EFT models to Horndeski one.  
The code will then internally set the behaviour of the EFT functions $\gamma_4$, $\gamma_5$, $\gamma_6$ according to eq.~(\ref{Horndeski_condition}) in order to cancel high order derivatives, at this point choices made for these functions will be ignored (see Section~\ref{pureHorndeski} for more details).
After setting these flags the user has to define the values of the EFT model parameters for the chosen model. Every other value of parameter and flag which do not concern the chosen model is automatically ignored.
\item The number $\texttt{EFTflag}=2$  corresponds to the implementation of alternative model-independent parametrizations in terms of EFT functions.   A lot of alternative parametrizations already present in literature can be completely described by using the versatility of the EFT approach allowing to preserve all the advantages of EFTCAMB. An alternative parametrization can be chosen by changing the flag $\texttt{AltParEFTmodel}$. See Section~\ref{Alternative} for details.
\item The number $\texttt{EFTflag}=3$  corresponds to the designer \textit{mapping} EFT procedure. Also in this case the \texttt{EFTwDE} flag controls the background expansion history which works as in the previous case.\\
For the \textit{mapping} case the user can investigate a particular DE/MG model once the matching with the EFT functions is provided and the background evolution has been implemented in the EFT code. \\
The model selection flag for the \textit{mapping} EFT procedure is \texttt{DesignerEFTmodel}.
Various models are already included in the code and their implementation details are presented in Section~\ref{Sec:MatchingEFT}. 
\item The number $\texttt{EFTflag}=4$  corresponds to the full \textit{mapping} EFT procedure. In this case the background expansion history is not set by a choice of $w_{\rm DE}$ and a model has to be fully specified. The code will then solve the background equations for the given model to map it into the EFT framework. Low energy Ho\v rava gravity has been included as the first example of the implementation of full mapping models. More models will be gradually filled in the near future. See Section~\ref{Fullmapping} for details.
\end{itemize}
\begin{figure}[!tp]
\begin{center}
 \includegraphics[width=\textwidth]{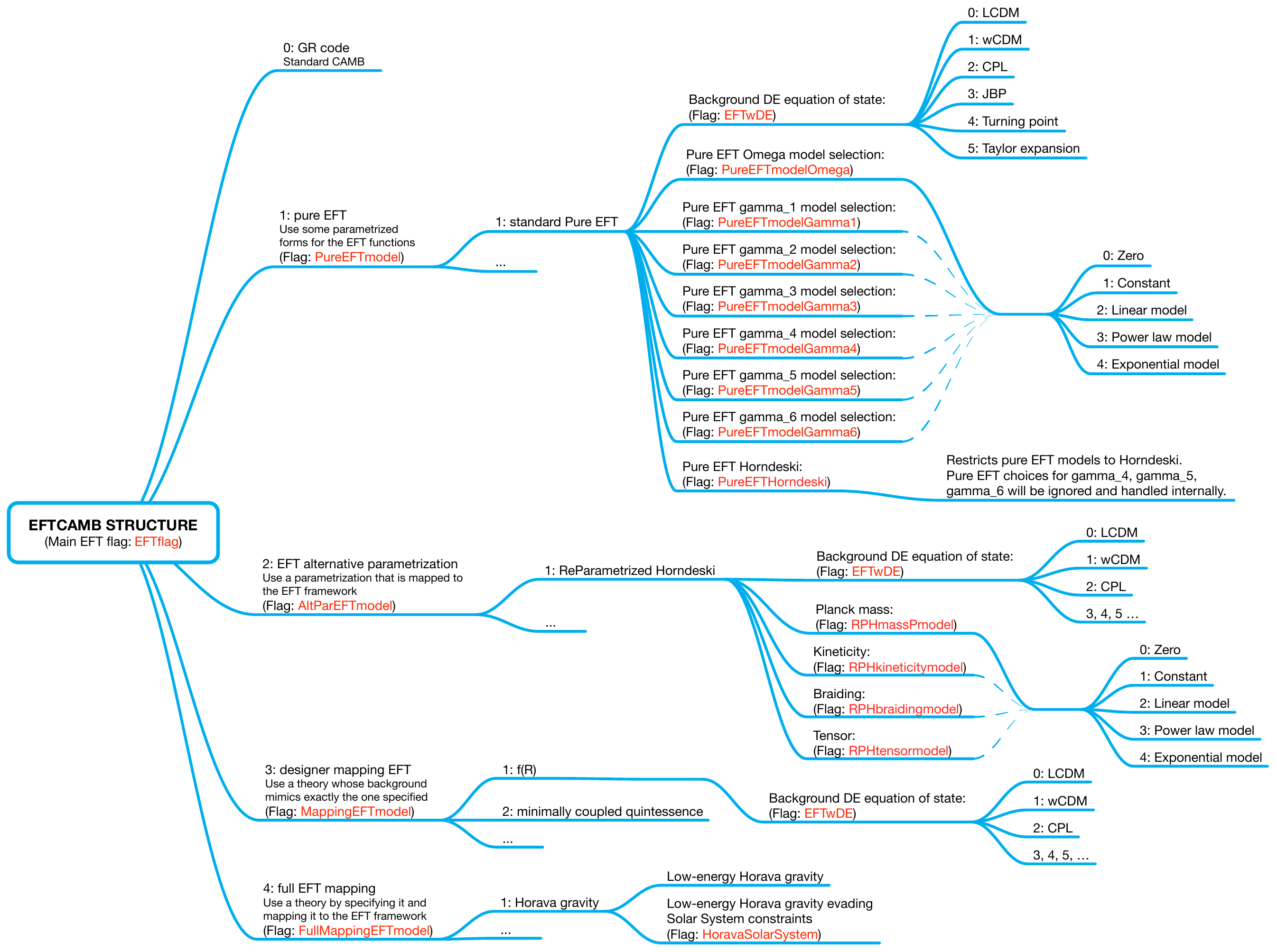}
 \caption{\label{Fig:StructureFlowChart} Flowchart of the structure of EFTCAMB: blue lines correspond to flags that are already present in the code.} 
\end{center}
\end{figure}
In addition EFTCAMB takes advantage of the feedback mechanism of CAMB with the following modifications:
\begin{itemize}
\item \texttt{feedback level=0} no feedback from EFTCAMB nor from EFTCosmoMC;
\item \texttt{feedback level=1} basic feedback, no feedback from EFTCAMB when called from EFTCosmoMC;
\item \texttt{feedback level=2} advanced feedback, no feedback from EFTCAMB when called from EFTCosmoMC;
\item \texttt{feedback level=3} debug feedback also when EFTCAMB is called from EFTCosmoMC;
\end{itemize}

\section{The structure of the modification}\label{Sec:ModStruct}
In order to implement the EFT formalism in the CAMB and CosmoMC codes we had to modify several files. \\
The complete, automatic, code documentation is available at \url{https://eftcamb.github.io/EFTCAMB/}. \\
To further help the user in understanding our part of code and/or applying the EFT modification to an already modified version of CAMB/CosmoMC we enclosed every modification that we made inside the following commented code lines:
\begin{verbatim}
! EFTCAMB MOD START
...
! EFTCAMB MOD END
\end{verbatim}
for the CAMB part and:
\begin{verbatim}
! EFTCOSMOMC MOD START
...
! EFTCOSMOMC MOD END
\end{verbatim}
for the CosmoMC part.\\

We also provide a developers version of the code at~\url{https://github.com/EFTCAMB/EFTCAMB}.
The tools provided by Github to visualize the history of the modifications of the code will, from now on, supersede the guides to the EFTCAMB/EFTCosmoMC modifications.

The step by step guide to the EFTCAMB modification v1.0 and v1.1 and v.2.0 will still be available at~\url{http://www.eftcamb.org/codes/guide_EFTCAMB.html} and the  EFTCosmoMC one at~\url{http://www.eftcamb.org/codes/guide_EFTCosmoMC.html}.

\section{Implementation of the modified equations}\label{Sec:ModEquations}
The implementation of the background in the code is described at length in~\cite{Hu:2013twa}. Here we shall review some of the more technical aspects and reproduce the equations in the form in which they enter the code. 

\subsection{General EFT Background} \label{SubSec:FullEFTBackground}
Starting from the general expressions for the cosmological background in EFT we can write the expansion history as a function of the EFT functions directly.
This results in:
\begin{align}
\hub^2 =& \frac{1}{1+\Omega +a\Omega'} \bigg[ \frac{1}{3} \frac{a^2\rho_{m,\nu}}{m_0^2} +\frac{2}{3}\frac{ca^2}{m_0^2}  -\frac{1}{3} \frac{\Lambda a^2}{m_0^2}\bigg] \nonumber \\
\dot{\hub} =& \frac{1}{1+\Omega+\frac{1}{2}a\Omega'} \bigg[ -\frac{1}{2}\left( 1+\Omega+2a\Omega' +a^2\Omega'' \right)\hub^2 -\frac{1}{2}\frac{P_{m,\nu}a^2}{m_0^2} -\frac{1}{2}\frac{\Lambda a^2}{m_0^2} \bigg] \nonumber \\
\ddot{\hub} =& \frac{1}{1+\Omega +\frac{1}{2}a\Omega' } \bigg[ -\frac{1}{2}a\hub^3\left( 3\Omega' +4a\Omega'' +a^2\Omega'''\right) -\hub\dot{\hub}\left( 1+\Omega +\frac{7}{2}a\Omega' +\frac{3}{2}a^2\Omega'' \right) \nonumber \\
&-\frac{1}{2}\left( \frac{\dot{P}_{m,\nu}a^2}{m_0^2} +2\hub \frac{P_{m,\nu}a^2}{m_0^2}\right) -\frac{1}{2}\left( \frac{\dot{\Lambda}a^2}{m_0^2} +2\hub \frac{\Lambda a^2}{m_0^2}\right) \bigg]
\end{align}
%

\subsection{Designer Background} \label{SubSec:EFTDesignerBackground}
Given the high degree of freedom already at the level of background, and since the focus will be on the dynamics of linear perturbations, it is common to adopt a designer approach as described in~\cite{Gubitosi:2012hu,Bloomfield:2012ff}. First of all one writes the background equations as follows:
\begin{align}
\mathcal{H}^2 =& \frac{8\pi G}{3} a^2 (\rho_m + \rho_{\rm DE} + \rho_{\nu})  \, ,\nonumber\\
\dot{\mathcal{H}} =& -\frac{4\pi G}{3} a^2 \left(\rho_m + \rho_{\rm DE} + \rho_{\nu} + 3P_m + 3 P_{\rm DE} + 3 P_{\nu} \right) = -\frac{\mathcal{H}^2}{2} - \frac{8 \pi G a^2 P_{\rm tot}}{2} \, ,\nonumber\\
\ddot{\mathcal{H}} =& 8 \pi G a^2 \rho_m \mathcal{H}\left(\frac{1}{6}+ w_m+\frac{3}{2}w_m^2\right) +  8 \pi G a^2 \rho_{\rm DE}\mathcal{H}\left( \frac{1}{6}+w_{\rm DE}+\frac{3}{2}w_{\rm DE}^2-\frac{1}{2}aw_{\rm DE}^{\prime} \right) \nonumber \\
  & +8 \pi G a^2 \left( \frac{\hub}{6}\rho_{\nu} -\frac{\hub}{2}P_{\nu} -\frac{1}{2}\dot{P}_{\nu} \right)
\label{Eq:BackDesigner}
\end{align}
where the prime stands for derivative w.r.t. the scale factor $a$, $\{\rho_m,P_m\}$ are the energy density and pressure of  matter (e.g.  dark matter, radiation and massless neutrinos)  and $\{\rho_{\rm DE},P_{\rm DE}\}$ encode the contributions from the extra scalar field into the form of an energy density and pressure of dark energy. For the matter components, one has the following continuity equations, and corresponding solutions:
\begin{align}
& \dot{\rho}_{m} + 3 \mathcal{H}\left(\rho_{m} + P_{m} \right)=0 \, ,& & \rho_{m}=\frac{3 H_0^2}{8 \pi G} \,\Omega_{m}^0\, a^{-3(1+w_m)} \, , \nonumber\\
& \dot{\rho}_{\rm DE} + 3 \mathcal{H}\rho_{\rm DE}\left[1 + w_{\rm DE}(a) \right]=0\, , & &  \rho_{\rm DE}= \frac{3 H_0^2}{8 \pi G}\, \Omega_{\rm DE}^0 \, \exp \bigg[-3 \int_{1}^{a}\frac{(1+w_{\rm DE}(a))}{a} \,da \bigg] \, ,
\label{Eq:BackContinuity}
\end{align}
where $\Omega^0_{m,\rm DE}$ is the energy density parameter  today, respectively of matter sector and dark energy, and $H_0$ is the present time Hubble parameter. 
Finally, $\{\rho_{\nu}, P_{\nu} \}$ are the density and pressure contributions due to massive neutrinos. The equation of state of massive neutrinos has a complicated, time dependent expression, hence the code computes directly $\dot{P}_{\nu}$. As the treatment of massive neutrinos in EFTCAMB  is  exactly the same as CAMB we refer the user to~\cite{Ma:1995ey,cambnotes}. However, let us stress that  in EFTCAMB  the indirect interaction (via gravity) of massive neutrino and DE/MG sectors  has been consistently taken into account both at the background and perturbations level, see also~\cite{Hu:2014sea}. With this setup, \emph{a background is fixed specifying $w_{\rm DE}$}. 
We illustrate in the next Section~\ref{SubSec:DeEOS} the models that are currently  implemented in the code. After the expansion history has been chosen one can then determine $c,\Lambda$ in terms of $\hub$ and $\Omega(a)$; namely, combining  eq.~(\ref{Eq:BackDesigner}) and eq.~(\ref{Eq:BackContinuity}) with the EFT background eqs.~(12,13) in~\cite{Hu:2013twa} one has:
\begin{align}\label{Eq:DesignerCLambda}
\frac{c a^2}{m_0^2} =& \left(\mathcal{H}^2- \dot{\mathcal{H}}\right) \left(\Omega +\frac{a\Omega^\prime}{2} \right) -\frac{a^2\mathcal{H}^2}{2}\Omega^{\prime\prime} + \frac{1}{2}\frac{a^2 \rho_{\rm DE}}{m_0^2}(1+w_{\rm DE})\, , & &   [\text{Mpc}^{-2}] \\
 \frac{\Lambda a^2}{m_0^2} =& -\Omega\left(2\mathcal{\dot{H}} + \mathcal{H}^2\right) -a\Omega^\prime \left(2\mathcal{H}^2 + \dot{\mathcal{H}} \right) - a^2\mathcal{H}^2 \Omega^{\prime \prime} + w_{\rm DE}\frac{a^2\rho_{\rm DE}}{m_0^2}\, , & & [\text{Mpc}^{-2}]  \\
\frac{\dot{c} a^2}{m_0^2} =& \frac{\mathcal{H}}{2}\left(-3 \left(1+w_{\rm DE}\right)^2+ a w^\prime_{\rm DE} \right)\frac{\rho_{\rm DE} a^2}{m_0^2} -\Omega \left(\ddot{\mathcal{H}}-4 \mathcal{H} \dot{\mathcal{H}} + 2 \mathcal{H}^3 \right)  + \frac{a\Omega^\prime}{2} \left(-\ddot{\mathcal{H}} + \mathcal{H}\dot{\mathcal{H}} +\mathcal{H}^3  \right) \nonumber \\
 &  +\frac{1}{2}a^2 \mathcal{H} \Omega^{\prime \prime}\left(\mathcal{H}^2 -3 \dot{\mathcal{H}} \right) -\frac{1}{2} a^3 \mathcal{H}^3 \Omega^{\prime\prime\prime}\, , & & [\text{Mpc}^{-3}] \\
\frac{\dot{\Lambda} a^2}{m_0^2} =& -2\Omega\left(\ddot{\mathcal{H}}- \mathcal{H} \dot{\mathcal{H}} -\mathcal{H}^3 \right) -a\Omega^{\prime}\left(5\mathcal{H}\dot{\mathcal{H}}+\ddot{\mathcal{H}} -\mathcal{H}^3\right)  - a^2\Omega^{\prime\prime}\mathcal{H}\left(2\mathcal{H}^2+3\dot{\mathcal{H}}\right)-a^3\mathcal{H}^3\Omega^{\prime\prime\prime} \nonumber\\
        & + \frac{\rho_{\rm DE} a^2}{m_0^2} \mathcal{H}\bigg[a w^\prime_{\rm DE} -3 w_{\rm DE}(1+w_{\rm DE}) \bigg] \, .& & [\text{Mpc}^{-3}]
\end{align}

As discussed in the Sections~\ref{Sec:PureEFTmodels} and~\ref{Sec:MatchingEFT}, depending on whether one wants to implement a \emph{pure} or \emph{mapping} EFT model, the choice for $\Omega$ changes. After fixing the expansion history, in the former case one selects an ans$\ddot{\text{a}}$tz for $\Omega(a)$, while in the latter case one determines via the matching the $\Omega(a)$ corresponding to the chosen model. In this case one has to separately solve the background equations for the given model, which might be done with a model-specific designer approach. In fact, this is the methodology we adopt for $f(R)$ models (see Section~\ref{SubSec:DesignerFR}).

\subsubsection{Effective Dark Energy equation of state parametrizations}\label{SubSec:DeEOS}
Several  models for the background expansion history have been implemented in the code:
\begin{itemize}
\item[] -  The $\Lambda$CDM expansion history: 
\begin{align}
& w_{\rm DE}=-1 \,,\nonumber \\
& \rho_{\rm DE} = m_0^2 \Lambda \,;
\end{align}
\item[] -  The $w$CDM model:
\begin{align}
& w_{\rm DE} = w_0 = \text{const} \neq -1 \,, \nonumber \\
& \rho_{\rm DE} = 3 m_0^2 H_0^2 \Omega^0_{\rm DE} a^{-3(1+w_0)} \,;
\end{align}
In code notation: $w_0 = \texttt{EFTw0}$.
\item[] -  The CPL parametrization~\cite{Chevallier:2000qy,Linder:2002et}: 
\begin{align}
& w_{\rm DE}(a)=w_0 + w_a(1-a) \,,\nonumber \\
& \rho_{\rm DE} = 3 m_0^2 H_0^2 \Omega^0_{\rm DE} a^{-3(1+w_0+w_a)}\exp\left(-3 w_a (1-a) \right) \,;
\end{align}
where $w_0$ and $w_a$ are constant and indicate, respectively, the value and the derivative of $w_{\rm DE}$ today. \\
In code notation: $w_0 = \texttt{EFTw0}$ and $w_a = \texttt{EFTwa}$.
\item[] - The generalized Jassal-Bagla-Padmanabhan parametrization~\cite{Jassal:2004ej,Jassal:2006gf}:
\begin{align}
& w_{\rm DE} = w_0 +(1-a)a^{n-1} w_a \,, \nonumber \\
& \rho_{\rm DE} = 3 m_0^2 H_0^2 \Omega^0_{\rm DE} a^{-3(1+w_0)}\exp \left(\frac{3 w_a \left((a (n-1)-n) a^n+a\right)}{a (n-1) n}\right) \,;
\end{align}
where $w_0$ is the value of $w_{\rm DE}$ for $a=1$ and $n=1$, while $n$ encodes the time of maximum deviation from $w_0$ and $w_a$ the extent of this deviation. For $n=1$ this reduces to the usual CPL parametrization. \\
In code notation: $w_0 = \texttt{EFTw0}$, $w_a = \texttt{EFTwa}$ and $n = \texttt{EFTwn}$.
\item[] - The {\it turning point} parametrization~\cite{Hu:2014ega}: 
\begin{align}
& w_{\rm DE} = w_0 +w_a\left(a_t-a \right)^2 \,, \nonumber \\
&  \rho_{\rm DE} = 3 m_0^2 H_0^2 \Omega^0_{\rm DE} a^{-3 \left(1+w_0+a_t^2 w_a\right)}\exp \left(-\frac{3}{2}w_a(a-1)\left(1+a-4a_t \right)\right) \,,
\end{align}
here $w_0$ is $w_{DE}(a=a_t)$ where $a_t$ is the value of the scale factor at the turning point, and  $w_a$ is its time derivative. 
In code notation: $w_0 = \texttt{EFTw0}$, $w_a = \texttt{EFTwa}$ and $a_t = \texttt{EFTwat}$. 
\item[] - The Taylor expansion around $a=0$:
\begin{align}
& w_{\rm DE} = w_0 + w_a a + \frac{1}{2}w_2 a^2 +\frac{1}{6}w_3 a^3 \,, \nonumber \\
& \rho_{\rm DE} = 3 m_0^2 H_0^2 \Omega^0_{\rm DE} a^{-3(1+w_0)} \exp \left((1-a) \left(3w_a + \frac{3}{4}w_2(a+1) +\frac{1}{6}w_3(a^2+a+1) \right)
\right) \,,
\end{align}
 where $w_2$ and $w_3$ are respectively the 2nd and 3rd time derivatives of $w_{DE}$.
In code notation: $w_0 = \texttt{EFTw0}$, $w_a = \texttt{EFTwa}$, $w_2 = \texttt{EFTw2}$ and $w_3 = \texttt{EFTw3}$.
\item[] - User defined: the EFTCAMB code includes the possibility for the user to define his/her own DE equation of state parametrization and it will properly account for it in any calculation. Let us notice that this option can be safely chosen without any modification in the structure of the code if the parametrized form of $w_{\rm DE}$ is given as a function of the scale factor $a(\tau)$.
\end{itemize}
These definitions of $w_{\rm DE}$ are shared by the pure EFT and designer EFT modules and can be consistently used for both choices of model.
%

\subsection{Other Background Quantities} \label{SubSec:EFTBackgroundQ}

Finally, for the purposes of the code, it is useful to compute the following EFT dark fluid components that can be derived from eq.~(10) of~\cite{Hu:2013twa}:
\begin{align}\label{Eq:EFTbackgroundDensity}
\frac{\rho_Q a^2}{m_0^2} =& 2 \frac{c a^2}{m_0^2} -\frac{\Lambda a^2}{m_0^2} - 3a \mathcal{H}^2 \Omega^\prime \, , & & [\text{Mpc}^{-2}]   \\ 
\frac{P_Q a^2}{m_0^2} =& \frac{\Lambda a^2}{m_0^2} + a^2 \mathcal{H}^2\Omega^{\prime \prime} + a \dot{\mathcal{H}} \Omega^\prime + 2 a \mathcal{H}^2 \Omega^\prime \, , 	& & [\text{Mpc}^{-2}]  \\
\frac{\dot{\rho}_Q a^2}{m_0^2} =& -3 \mathcal{H}\left(\frac{\rho_Q a^2}{m_0^2}+ \frac{P_Q a^2}{m_0^2} \right) + 3 a \mathcal{H}^3 \Omega^\prime \, , & & [\text{Mpc}^{-3}]  \\
\frac{\dot{P_Q} a^2}{m_0^2} =&\frac{\dot{\Lambda}a^2}{m_0^2}+ a^3 \mathcal{H}^3\Omega^{\prime\prime\prime}+3a^2\mathcal{H}\dot{\mathcal{H}}\Omega^{\prime\prime}+a\Omega^{\prime}\ddot{\mathcal{H}}+3a\mathcal{H}\dot{\mathcal{H}}\Omega^{\prime}+2a^2\mathcal{H}^3\Omega^{\prime\prime}-2a\mathcal{H}^3\Omega^{\prime} \, .
              & &   [\text{Mpc}^{-3}]         
%
\end{align}
%

\subsection{Linear Scalar Perturbations in EFT: code notation} \label{SubSec:EFTPe}
In this Section we write the relevant equations that EFTCAMB uses~\cite{Ma:1995ey}. We write them in a compact notation that almost preserves the form of the standard equations simplifying both the comparison with the GR limit and the implementation in the code. \\
The dynamical equations that EFTCAMB evolves can be written\footnote{Notice that working in the Jordan frame ensures that the energy-momentum conservation equations 
will not change with respect to their GR form. For this reason the evolution equations for density and velocity are not reported here.} as:
\begin{align}
& A(\tau,k)\,\ddot{\pi} + B(\tau,k)\,\dot{\pi} + C(\tau)\,\pi + k^2\,D(\tau,k)\,\pi + H_0 E(\tau,k) =0 \label{Eq:PiFieldEquation} \,,\\
& k\dot{\eta} = \f{1}{X}\l[\frac{1}{1+\Omega} \frac{a^2(\rho_{m,\nu}+P_{m,\nu})}{m_0^2}\frac{v_{m,\nu}}{2} + \frac{k^2}{3H_0}F +(U-X)\f{\mathcal{Z}k^2}{3}\r]   \label{Eq:EtaEquation}\,,
\end{align}
while constraint equations take the form:
\begin{align}
\sigma =&\f{1}{X}\l[ \mathcal{Z}U + \frac{1}{1+\Omega}\frac{3}{2k^2}\frac{a^2(\rho_{m,\nu}+P_{m,\nu})}{m_0^2}v_{m,\nu} + \frac{{F}}{H_0}\r] \label{Eq:SigmaEquation}\,,\\
\dot{\sigma} =&\f{1}{X}\l[ -2\mathcal{H}\l[1+V\r]\sigma  + k \eta-\frac{1}{k}\frac{a^2P_{m,\nu}}{m_0^2}\frac{\Pi_{m,\nu}}{1+\Omega} +\f{N}{H_0}\r] \label{Eq:SigmadotEquation}\,,\\
\ddot{\sigma} =& \f{1}{X}\l[ -2\l(1+V \r)\l(\dot{\mathcal{H}}\sigma +\hub \dot{\sigma} \r) -2\hub \sigma \dot{V} +k \dot{\eta} +\frac{1}{k}\frac{a \mathcal{H} \Omega^\prime}{(1+\Omega)^2}\frac{a^2P_{m,\nu}}{m_0^2}\Pi_{m,\nu} -\frac{1}{k(1+\Omega)}\frac{d}{d\tau}\left(\frac{a^2P_{m,\nu}}{m_0^2}\Pi_{m,\nu} \right)-\dot{{X}}\dot{\sigma} +\f{\dot{N}}{H_0}\r] \label{Eq:SigmadotdotEquation}\,,\\
\mathcal{Z} =& \f{1}{G}\l[\mathcal{Q}\frac{k\eta}{\mathcal{H}} + \frac{1}{2\mathcal{H}(1+\Omega)k}\frac{a^2 \delta\rho_{m,\nu}}{m_0^2} + \frac{{L}}{kH_0} \r] \label{Eq:Zequation}\,,\\
\dot{\mathcal{Z}} =&\f{1}{U}\l[ -2 \mathcal{H} \l[1 + Y\r] \mathcal{Z} + k\eta -\frac{1}{1+\Omega}\frac{3}{2k}\frac{a^2\delta P_{m,\nu}}{m_0^2} - \frac{3}{2k(1+\Omega)}\frac{{M}}{H_0}\r] \,,\nonumber \\
 =& \f{1}{U}\l[-2\hub\mathcal{Z}\l(1+Y-\f{G}{2}\r) -\f{1}{2(1+\Omega)k}\f{a^2\delta\rho_{m,\nu}}{m_0^2} -\f{3}{2(1+\Omega)k}\f{a^2\delta P_{m,\nu}}{m_0^2} -\f{\hub L}{H_0 k} -\f{3}{2(1+\Omega)k}\f{M}{H_0} \r] \label{Eq:ZdotEquation}\,,
\end{align}
where $2k\mathcal{Z}\equiv \dot{h}$ and $2k\sigma_*\equiv\dot{h}+6\dot{\eta}$ are the standard CAMB variables. 
In these expressions we wrote the same prefactor, $X$, in eqs.~(\ref{Eq:SigmaEquation}) and~(\ref{Eq:SigmadotEquation}), and $U$, in eqs.~(\ref{Eq:SigmaEquation}) and~(\ref{Eq:ZdotEquation}), but we have to stress that they might be different if other second order EFT operators are considered.
In addition the last expression~(\ref{Eq:ZdotEquation}) has two forms: the first is the standard one while the second one is used when the CAMB code uses the RSA approximation.\\
At last, to compute the observable spectra we had to define two auxiliary quantities:
\begin{align}
\texttt{EFTISW} =& \,\ddot{\sigma} + k\dot{\eta} = \f{1}{X}\bigg[ -2\l[1+V \r]\l(\dot{\mathcal{H}}\sigma +\hub \dot{\sigma} \r) -2\hub \sigma \dot{V} +\f{\l(1+X \r)}{2(1+\Omega)X}\frac{a^2(\rho_{m,\nu}+P_{m,\nu})}{m_0^2}v_{m,\nu}  \nonumber\\
	& \hspace{2.3cm} +\frac{1}{k}\frac{a \mathcal{H} \Omega^\prime}{(1+\Omega)^2}\frac{a^2P_{m,\nu}}{m_0^2}\Pi -\frac{1}{k(1+\Omega)}\frac{d}{d\tau}\left(\frac{a^2P_{m,\nu}}{m_0^2}\Pi_{m,\nu} \right) +\f{(1+X)k^2}{3 H_0 X}F  \nonumber \\
	& \hspace{2.3cm} +\f{(1+X)k^2}{3 X}\mathcal{Z}(U-X) -\dot{X}\dot{\sigma} +\f{\dot{N}}{H_0} \bigg] \label{Eq:TTSource}\,,\\
\texttt{EFTLensing} =& \,\dot{\sigma} + k\eta = \f{1}{X}\l[ -2\mathcal{H}\l(1 + V\r)\sigma +(1+ X) k \eta -\frac{1}{k (1+ \Omega)}\frac{a^2 P_{m,\nu}}{m_0^2} \Pi_{m,\nu} +\f{N}{H_0} \r] \label{Eq:LensingSource}\,.
\end{align}
The coefficients for the $\pi$ field equation, once a complete de-mixing is achieved, can not be divided into contributions due to one operator at a time so we write here their full form:
\begin{align}
A =& \f{c a^2}{m_0^2}+2 a^2 H_0^2 \gamma_1 +\f{3}{2}a^2 \f{\l(\hub \Omega' + H_0\gamma_2 \r)^2}{2(1+\Omega) + 3\gamma_3 +\gamma_4} + 4 \gamma_6 k^2 \,, \\
& \nonumber \\
B =& \f{\dot{c}a^2}{m_0^2} +4\hub \f{c a^2}{m_0^2} +8 a^2\hub H_0^2\l(\gamma_1 +a\f{\gamma_1^\prime}{4}\r) +4k^2\hub\l(2\gamma_6 +a\gamma_6^\prime \r) + ak^2 \f{\gamma_4+2\gamma_5}{2(1+\Omega) -2\gamma_4}\l(\hub\Omega' +H_0\gamma_2\r) \nonumber\\
& -a\f{\hub \Omega' + H_0\gamma_2}{4(1+\Omega)+6\gamma_3 +2\gamma_4}\bigg[ -3\f{a^2(\rho_{Q}+P_Q)}{m_0^2} -3a\hub^2\Omega'\l( 4+ \f{\dot{\hub}}{\hub^2} +a\f{\Omega''}{\Omega'}\r) -3a\hub H_0 \l(4\gamma_2+a\gamma_2^\prime\r) \nonumber\\
& +\l(9 \gamma_3+3\gamma_4 \r)\l(\dot{\hub} -\hub^2\r) +k^2\l(-3\gamma_3-\gamma_4 +4\gamma_5 \r)\bigg]  \nonumber \\
& +\f{1}{1+\Omega+2\gamma_5}\l( a\hub\Omega' +2\hub\l(\gamma_5+a\gamma_5^\prime\r) -(1+\Omega)\f{a\hub\Omega' +aH_0\gamma_2}{2(1+\Omega) +3\gamma_3+\gamma_4}\r)\cdot \nonumber\\
&\cdot\l[-\f{a^2c}{m_0^2}+\f{3}{2}a\hub^2\Omega'-2a^2H_0^2\gamma_1-4\gamma_6k^2 +\f{3}{2}a\hub H_0\gamma_2 \r] \, , \\
& \nonumber \\
C =&  \hub\f{\dot{c}a^2}{m_0^2} +\l(6\hub^2-2\dot{\hub}\r)\f{ca^2}{m_0^2}  +\f{3}{2}a\hub\Omega'\l(\ddot{\hub} -2\hub^3\r) +6\hub^2H_0 ^2\gamma_1a^2 + 2a^2\dot{\hub} H_0^2\gamma_1 +8a^3\hub^2 H_0^2\f{\gamma_1^\prime}{4} \nonumber\\
& +\f{3}{2}\l(\dot{\hub} -\hub^2\r)^2\l(\gamma_4 +3\gamma_3\r) +\f{9}{2}\hub H_0 a\l(\dot{\hub}-\hub^2\r)\l(\gamma_2+a\f{\gamma_2^\prime}{3}\r) +\f{a}{2}H_0\gamma_2\l(3\ddot{\hub}-12\dot{\hub}\hub +6\hub^3\r) \nonumber \\
& -a\f{\hub \Omega' + H_0\gamma_2}{4(1+\Omega)+6\gamma_3+2\gamma_4}\bigg[ -3\f{a^2\dot{P}_Q}{m_0^2}-3\hub \l(\f{a^2\rho_Q}{m_0^2} +\f{a^2 P_Q}{m_0^2}\r) -3a\hub^3\l(a\Omega'' +6\Omega' +2\f{\dot{\hub}}{\hub^2}\Omega' \r) \nonumber\\
& +3\l(\ddot{\hub}-2\hub \dot{\hub}\r)\l(\gamma_4 +3\gamma_3\r) +6\hub\l(\dot{\hub}-\hub^2\r)\l(3\gamma_3+3a\f{\gamma_3^\prime}{2} +\gamma_4+a\f{\gamma_4^\prime}{2}\r) \nonumber \\
& -3a H_0\l(3\hub^2\gamma_2+\dot{\hub}\gamma_2 +a\hub^2\gamma_2^\prime\r) \bigg] \nonumber \\
& +\f{1}{1+\Omega+2\gamma_5}\l( a\hub\Omega' +2\hub\l(\gamma_5+a\gamma_5^\prime\r) -(1+\Omega)\f{a\hub\Omega' +aH_0\gamma_2}{2(1+\Omega) +3\gamma_3+\gamma_4}\r)\cdot \nonumber\\
&\cdot\bigg[ -\f{1}{2}\f{a^2\dot{\rho}_Q}{m_0^2} -\hub\f{a^2c}{m_0^2} +\f{3}{2}a\hub\Omega'\l(3\hub^2 -\dot{\hub}\r) -2a^2\hub H_0^2\gamma_1 -\f{3}{2}aH_0\gamma_2\l(\dot{\hub}-2\hub^2\r) -3\hub\l(\dot{\hub}-\hub^2\r)\l(\f{3}{2}\gamma_3 + \f{\gamma_4}{2}\r)\bigg] \, ,\nonumber\\
& \nonumber \\
D =& \f{ca^2}{m_0^2}-\f{1}{2}a\hub H_0\l( \gamma_2+a\gamma_2^\prime\r) +\l(\hub^2-\dot{\hub}\r)\l(3\gamma_3+\gamma_4\r) +4\l( \dot{\hub}\gamma_6+\hub^2\gamma_6 +a\hub^2\gamma_6^\prime\r) +2\l(\dot{\hub}\gamma_5+a\hub^2\gamma_5^\prime \r) \nonumber\\
& -a\f{\hub \Omega' + H_0\gamma_2}{4(1+\Omega)+6\gamma_3+2\gamma_4}\bigg[ -2a\hub\Omega' +4\hub\gamma_5 -2\hub\l(3\gamma_3+3a\f{\gamma_3^\prime}{2}+ \gamma_4 +a\f{\gamma_4^\prime}{2} \r)\bigg] \nonumber\\
& +\f{1}{1+\Omega+2\gamma_5}\l( a\hub\Omega' +2\hub\l(\gamma_5+a\gamma_5^\prime\r) -(1+\Omega)\f{a\hub\Omega' +aH_0\gamma_2}{2(1+\Omega) +3\gamma_3+\gamma_4}\r)\cdot \nonumber\\
& \cdot \bigg[ \f{1}{2}a\hub \Omega' -2\hub\gamma_5+\f{1}{2}aH_0 \gamma_2+\f{3}{2}\hub \gamma_3 +\hub\f{\gamma_4}{2} -4\hub \gamma_6 \bigg] \nonumber \\
& +\f{\gamma_4 +2\gamma_5}{2(1+\Omega) -2\gamma_4}\bigg[\f{a^2\l(\rho_Q+ P_Q \r)}{m_0^2} +a\hub^2\Omega' -\gamma_4\l(\dot{\hub}-\hub^2\r) +a\hub H_0\gamma_2+3\gamma_3\l(\hub^2 -\dot{\hub} \r)\bigg] \nonumber \\
& +k^2\bigg[\f{\gamma_3}{2} +\f{\gamma_4}{2} +\f{\gamma_4 +2\gamma_5}{2(1+\Omega)-2\gamma_4}\l(\gamma_3+\gamma_4 \r) \bigg] \, ,\nonumber \\
& \nonumber \\
E =& \bigg\{ \f{ca^2}{m_0^2} -\f{3}{2}a\hub^2\Omega' -\f{1}{2}a\hub H_0\l(2\gamma_2+a\gamma_2^\prime\r) +\f{1}{2}\gamma_3\l(k^2 -3\dot{\hub} + 3\hub^2\r) +\f{1}{2}\gamma_4\l(k^2 -\dot{\hub} +\hub^2 \r) \nonumber \\
& -a\f{\hub \Omega' + H_0\gamma_2}{4(1+\Omega)+6\gamma_3+2\gamma_4}\bigg[ -2\hub\l(a\Omega' +2(1+\Omega)\r) -2\hub\l(3\gamma_3 +3 a\f{\gamma_3^\prime}{2} +\gamma_4 +a\f{\gamma_4^\prime}{2}\r) \bigg] \nonumber \\
& +\f{1}{1+\Omega+2\gamma_5}\l( a\hub\Omega' +2\hub\l(\gamma_5+a\gamma_5^\prime\r) -(1+\Omega)\f{a\hub\Omega' +aH_0\gamma_2}{2(1+\Omega) +3\gamma_3+\gamma_4}\r)\cdot \nonumber\\
&\cdot \bigg[ \hub\l(1+\Omega +\f{a\Omega'}{2}\r) +\f{1}{2}aH_0\gamma_2+\f{3}{2}\hub\gamma_3+\f{\hub\gamma_4}{2} \bigg] +\f{\gamma_4 +2\gamma_5}{2(1+\Omega) -2\gamma_4}k^2\l(\gamma_4 +\gamma_3\r) \bigg\} k\mathcal{Z} \nonumber \\
& +3a\f{\hub \Omega' + H_0\gamma_2}{4(1+\Omega)+6\gamma_3 +2\gamma_4}	\l(\f{a^2 \delta P_{m,\nu}}{m_0^2} \r) +\f{\gamma_4 +2\gamma_5}{2(1+\Omega) -2\gamma_4}k\l(\f{a^2\l(\rho_{m,\nu} +P_{m,\nu} \r)v_{m,\nu}}{m_0^2} \r) \nonumber \\
& -\f{1}{2}\f{1}{1+\Omega+2\gamma_5}\l( a\hub\Omega' +2\hub\l(\gamma_5+a\gamma_5^\prime\r) -(1+\Omega)\f{a\hub\Omega' +aH_0\gamma_2}{2(1+\Omega) +3\gamma_3+\gamma_4}\r)\l(\f{a^2\delta\rho_{m,\nu}}{m_0^2} \r) \, .\label{EEFT}
\end{align}
On the other hand the non-zero contributions to be added to the Einstein equations can be written for each operator separately and are listed in the following as $\Delta F,\Delta G, \Delta N,...$ respectively.\\
We adopt the following convention: $F=\sum \Delta F$ and the same applies to all the other terms.
\begin{itemize}
\item[] \underline{Background operators}:\\
{\small
\begin{align} \label{Eq:BackgroundOp}
\Delta F =& \frac{3}{2k(1+\Omega)}\bigg[\frac{(\rho_Q+P_Q)a^2}{m_0^2}\pi + a \mathcal{H} \Omega^\prime \left( \dot{\pi} + \mathcal{H} \pi \right) \bigg] \, , & & [\text{Mpc}^{-1}] \,, \nonumber \\
\Delta G =&  \left(1+ \frac{a\Omega^\prime}{2(1+\Omega)} \right)\,, \, & & [\,\,] \, \nonumber \\
\Delta L =& -\frac{3}{2}\frac{a\Omega^\prime}{1+\Omega}(3 \mathcal{H}^2 - \dot{\mathcal{H}})\pi -\frac{3}{2}\frac{a\Omega^\prime}{1+\Omega}\mathcal{H}\dot{\pi} -\frac{1}{2}\frac{a\Omega^\prime}{1+\Omega}k^2\pi + \frac{\pi}{2\mathcal{H}(1+\Omega)}\frac{a^2\dot{\rho}_Q}{m_0^2} + \frac{\dot{\pi}+ \mathcal{H}\pi}{\mathcal{H}(1+\Omega)}\frac{a^2c}{m_0^2}  \, , & & [\text{Mpc}^{-2}] \, \nonumber \\
\Delta M =& \frac{\dot{P}_Qa^2}{m_0^2}\pi + \frac{(\rho_Q + P_Q)a^2}{m_0^2}\left(\dot{\pi} + \mathcal{H} \pi \right) + a\mathcal{H}\Omega^\prime \left[ \ddot{\pi}+\left(\frac{\dot{\hub}}{\hub}+4\hub+a\hub\frac{\Omega^{\prime\prime}}{\Omega^{\prime}}\right)\dot{\pi} \right. \nonumber \\ 
&\left. + \left(2\dot{\hub}+6\hub^2+a\hub^2\frac{\Omega^{\prime\prime}}{\Omega^{\prime}}+\frac{2}{3}k^2\right) \pi\right] \, , & & [\text{Mpc}^{-3}] \, \nonumber \\
\Delta N =& -\f{a\hub\Omega'}{1+\Omega}k\pi \,, \hspace{1cm} \dot{N} =-\f{a\dot{\hub}\Omega'}{1+\Omega}k\pi  -\f{a\hub\Omega'}{1+\Omega}k\dot{\pi} -\f{a\hub^2}{(1+\Omega)}\l[\Omega' +a\Omega'' -\f{a\Omega'^{\,2}}{1+\Omega} \r]k\pi  \, , & & [\text{Mpc}^{-2}, \text{Mpc}^{-3}] \, \nonumber \\
\Delta X=&1 \,, \hspace{1cm} \Delta \dot{X}=0  \, , & & [\,\, , \,\,] \, \nonumber \\
\Delta Y=&  \frac{a\Omega^\prime}{2(1+\Omega)}  \, , & & [\,\, ] \, \nonumber \\
\Delta U=&1, \, & & [\,\, ] \, \nonumber \\
\Delta V=&\f{1}{2}\f{a\Omega'}{1+\Omega} \,, \hspace{1cm} \Delta \dot{V}=\f{a\hub}{2(1+\Omega)}\l[\Omega' +a\Omega'' -\f{a\Omega'^{\,2}}{1+\Omega} \r] \, , & & [\,\, , \text{Mpc}^{-1}]\nonumber \\ \,
\Delta\mathcal{Q}=& 1\,, & & [\,\, ]
\end{align}}
\item[] \underline{$(\delta g^{00})^2$}:\\
{\small
\begin{align}\label{Eq:ContribGamma1}
& \Delta {{ L}}= \f{2a^2H_0^2\gamma_1}{\hub \l(1 + \Omega \r)}\l(\dot{\pi}+\hub\pi\r)\, .& & [\text{Mpc}^{-2}] \,
\end{align}}
\item[] 
\underline{$\delta g^{00}\delta K^\mu_\mu$}:
{\small
\begin{align} \label{Eq:ContribGamma2}
&\Delta{ G}=\f{a H_0 \gamma_2}{2\hub(1+\Omega)} \, , & & [\,\,] \,\nonumber \\
&\Delta{ F}=\f{3}{2}aH_0\gamma_2\f{\dot{\pi}+\hub\pi}{k(1+\Omega)}\, , & &  [\text{Mpc}^{-1}] \,\nonumber\\
&\Delta{ L}=\f{3}{2} \f{a H_0\gamma_2}{1+\Omega}\left[\l(\f{\dot{\hub}}{\hub}-2\hub-\f{k^2}{3\hub}\r)\pi-\dot{\pi}\r] \, ,& &  [\text{Mpc}^{-2}] \,\nonumber\\
&\Delta{ M}= aH_0\l[\gamma_2\ddot{\pi}+\l(4\gamma_2+a\gamma_2^\prime\r)\hub\dot{\pi}+\l(3\hub^2\gamma_2+\dot{\hub}\gamma_2+a\hub^2\gamma_2^\prime\r)\pi\r]\, . & &  [\text{Mpc}^{-3}] \,
\end{align}}
\item[] \underline{$(\delta K)^2$}:\\
{\small
\begin{align} \label{Eq:ContribGamma3}
&\Delta{ G}=\f{3}{2}\f{\gamma_3}{1+\Omega}\, , & &  [\,\,]  \,\nonumber \\
&\Delta{ F}=+\f{3}{2}\f{\gamma_3}{1+\Omega}\l[k-3\f{\dot{\hub}-\hub^2}{k}\r]\pi \, , & &  [\text{Mpc}^{-1}] \,\nonumber\\
&\Delta{ L}= -\f{3}{2}\f{\gamma_3}{1+\Omega}\l[k^2-3\l(\dot{\hub}-\hub^2\r)\r]\pi \, ,& &  [\text{Mpc}^{-2}] \,\nonumber\\
&\Delta{ M}= \gamma_3\l(3\hub^2-3\dot{\hub}+k^2\r)\dot{\pi}+\gamma_3 \l(6\hub^3-3\ddot{\hub} \r)\pi +2\hub k^2\pi\l(\gamma_3+a\f{\gamma_3^\prime}{2}\r) -6a\hub(\dot{\hub}-\hub^2)\f{\gamma_3^\prime}{2}\pi \, , & &  [\text{Mpc}^{-3}] \,\nonumber\\
& \Delta{ Y}= \f{3}{2(1+\Omega)}\l( \gamma_3+a\f{\gamma_3^\prime}{2}\r) \, , & &  [\,\,] \,\nonumber\\
&\Delta{ U}=\f{3}{2}\f{\gamma_3}{1+\Omega} \, . & &  [\,\,] \,
\end{align}}
\item[] \underline{$\delta K^\mu_\nu\delta K^\nu_\mu$}:\\
{\small
\begin{align} \label{Eq:ContribGamma4}
\Delta{ G} =& \f{\gamma_4}{2(1+\Omega)} \, , & &  [\,\,]  \, \nonumber \\
\Delta{ F} =& +\f{3\gamma_4}{2(1+\Omega)}k\pi-\f{3\gamma_4 }{2(1+\Omega)}\f{\dot{\hub}-\hub^2}{k}\pi \, ,& &  [\text{Mpc}^{-1}] \,\nonumber\\
\Delta{ X} =& -\f{\gamma_4}{1+\Omega} \,,\hspace{1cm} \Delta\dot{X}=-\f{a\hub}{1+\Omega}\l[\gamma_4^\prime-\f{\gamma_4\Omega'}{1+\Omega} \r] \, , & &  [\,\,\,, \text{Mpc}^{-1}]  \, \nonumber \\
\Delta{ L} =& \f{3\gamma_4}{2(1+\Omega)}\l(\dot{\hub}-\hub^2-\f{k^2}{3}\r)\pi \, , & &  [\text{Mpc}^{-2}] \,\nonumber\\
\Delta{ N} =& \f{2 \hub}{1+\Omega}\l(\gamma_4+a\f{\gamma_4^\prime}{2} \r)k\pi + \f{\gamma_4}{1+\Omega}k\dot{\pi}\, , & &  [\text{Mpc}^{-2}] \,\nonumber\\
\Delta\dot{N} =& \f{\gamma_4k\ddot{\pi}}{1+\Omega} +\f{a\hub k\dot{\pi}}{1+\Omega}\l[\gamma_4^\prime-\f{\gamma_4\Omega'}{1+\Omega} \r] +\f{2k}{1+\Omega}\l(\gamma_4+a\f{\gamma_4^\prime}{2} \r)\l( \dot{\hub}\pi + \hub \dot{\pi} \r) \nonumber\\
& +\f{2a\hub^2k\pi}{1+\Omega}\l[a \f{\gamma_4^{\prime\prime}}{2}+3\f{\gamma_4^\prime}{2}-\f{\Omega'}{1+\Omega}\l(\gamma_4+a\f{\gamma_4^\prime}{2} \r) \r] \, ,& &  [\text{Mpc}^{-3}]  \, \nonumber \\
\Delta{ V} =& -\f{1}{1+\Omega}\l(\gamma_4+a\f{\gamma_4^\prime}{2} \r) \, ,& &  [\,\,]  \, \nonumber \\
\Delta\dot{V} =& -\f{a\hub}{1+\Omega}\l[a\f{\gamma_4^{\prime\prime}}{2} +3\f{\gamma_4^\prime}{2}-\f{\Omega'}{1+\Omega}\l(\gamma_4+a\f{\gamma_4^\prime}{2} \r) \r] \, , & &  [\text{Mpc}^{-1}]  \, \nonumber \\
\Delta{ Y} =& \f{1}{2(1+\Omega)}\l(\gamma_4+a\f{\gamma_4^\prime}{2} \r) \, ,& &  [\,\,]  \ \nonumber \\
\Delta{ M} =& -\gamma_4\l(\dot{\hub}-\hub^2-\f{k^2}{3}\r)\dot{\pi} -2\hub\l(\gamma_4+a\f{\gamma_4^\prime}{2} \r)\l(\dot{\hub}-\hub^2-\f{k^2}{3} \r)\pi -\gamma_4\l(\ddot{\hub}-2\hub\dot{\hub}\r)\pi  \, ,& &  [\text{Mpc}^{-3}] \,\nonumber\\
\Delta{ U} =& \f{\gamma_4}{2(1+\Omega)} \, . & &  [\,\,]  \,
\end{align}}
\item[] \underline{$\delta g^{00}\delta R^{(3)}$}:\\
{\small
\begin{align}\label{Eq:ContribGamma5}
&\Delta{ M}=-\f{4\gamma_5k^2}{3}\l(\dot{\pi}+\hub\pi \r) \, , & &  [\text{Mpc}^{-3}] \, \nonumber\\
&\Delta{ N}=\f{2\gamma_5k}{1+\Omega}\left(\dot{\pi}+\hub\pi\right) \, , & &  [\text{Mpc}^{-2}] \, \nonumber\\
&\Delta\dot{N} = \f{2\gamma_5k}{1+\Omega}\l(\ddot{\pi}+\hub\dot{\pi}+\dot{\hub}\pi\r)+\f{2ak\hub}{1+\Omega}\l(\dot{\pi}+\hub\pi\r)\l[ \gamma_5^\prime-\f{\gamma_5\Omega'}{1+\Omega}\r]  \, ,& &  [\text{Mpc}^{-3}] \,\nonumber\\
&\Delta\mathcal{Q}= \f{2\gamma_5}{1+\Omega}\,, & & [\,\, ] \nonumber \\
&\Delta L= \f{2\gamma_5}{1+\Omega}k^2 \pi\,.& &  [\text{Mpc}^{-2}]
\end{align}}
\item[] \underline{ $\rm(g^{\mu\nu}+n^\mu n^\nu)\partial_\mu\delta g^{00}\partial_\nu \delta g^{00}$}:\\
{\small
\begin{align}\label{Eq:ContribGamma6}
&\Delta{ L}= \f{4\gamma_6k^2}{\hub(1+\Omega)}\l(\dot{\pi}+\hub\pi\r) \, . & &  [\text{Mpc}^{-2}] \,
\end{align}}
\end{itemize}
%
\subsection{Linear Tensor Perturbations in EFT: code notation}\label{SubSec:TensorModes}
The tensor component of the B-mode polarization of the CMB can be used to further constrain modifications of gravity~\cite{Amendola:2014wma,Raveri:2014eea}.
In this Section we write the relevant equation that EFTCAMB uses to build the tensor component of the CMB spectra.
Since we are working in the Jordan frame only the propagation equation for tensor perturbations needs to be modified into:
\begin{align} \label{Eq:TensorEquation}
A_{T}(\tau) \ddot{h}_{ij} +B_{T}(\tau) \dot{h}_{ij} +D_{T}(\tau) k^2 h_{ij}+ E_{Tij} = 0,
\end{align}
where:
\begin{align}
A_{T} =& 1+\Omega -\gamma_4 \,,\nonumber \\
B_{T} =& 2\hub \left(1 + \Omega -\gamma_4 +\frac{a\Omega'}{2} -a \f{\gamma_4^\prime}{2}	 \right)\,, \nonumber \\
D_{T} =& 1+\Omega\,,\nonumber \\
E_{Tij}=& \frac{a^2}{m_0^2} \delta T _{ij} \,,
\end{align}
and $\delta T_{ij}$ contains the neutrinos and photons contribution to the tensor component of anisotropic stress.
%
\subsection{Viability conditions}\label{SubSec:ViabilityConditions}
In this Section we list the viability priors that EFTCAMB naturally enforces on the EFT functions in order to ensure that the theory under consideration is stable. We separate these conditions in a set of physical ones and a set of mathematical ones, as  described in the following. The full set of physical conditions for the general EFT action is a very complex matter that is the focus of our ongoing work. Here we report a preliminary version of them that contains the ghost and gradient stability for the subset of the EFT corresponding to GLPV theories~\cite{Gleyzes:2014dya}, for which the mixing between the matter and gravity perturbations has been considered~\cite{Kase:2014cwa}. The full class of Horndeski models is included in these theories. 
Specifically, when $\gamma_3=-\gamma_4$ and $\gamma_6=0$ (i.e. we consider GLPV  theories) the kinetic and gradient stability conditions reduce respectively to
\ba
&&\tilde{W}_2\l[4\tilde{W}_1\tilde{W}_2-\tilde{W}_3^2\r]>0,\\
&&\tilde{W}_0\tilde{W}_3^2+a\hub\l(\tilde{W}_2\tilde{W}_3\tilde{W}_6^\prime+\tilde{W}_6\tilde{W}_3\tilde{W}_2^\prime-\tilde{W}_6\tilde{W}_2\tilde{W}_3^\prime\r)+2\hub\tilde{W}_3\tilde{W}_2\tilde{W}_6>\f{9}{2}\tilde{W}_6^2\f{a^2}{m_0^2}(\rho_m+P_m),
\ea
where
\ba
&&\tilde{W}_0=-(1+\Omega), \\
&&\tilde{W}_1=\f{ca^2}{m_0^2}+2H_0^2a^2\gamma_1-3\hub^2(1+\Omega)-3a\hub^2\Omega^\prime+3\hub^2\gamma_4-3aH_0\hub\gamma_2, \\
&&\tilde{W}_2=-3[(1+\Omega)-\gamma_4], \\
&&\tilde{W}_3=6\hub(1+\Omega)+3a\hub\Omega^\prime-6\hub\gamma_4+3aH_0\gamma_2,\\
&&\tilde{W}_6=-4\l[\f{1}{2}(1+\Omega)+\gamma_5\r],\\
&&\tilde{W}_2^\prime=-3[\Omega^\prime-\gamma_4^\prime]\\
&&\tilde{W}_3^\prime= 6\f{\dot{\hub}}{a\hub}(1+\Omega)+9\hub\Omega^\prime+3\f{\dot{\hub}}{\hub}\Omega^\prime+3a\hub\Omega^{\prime\prime}-6\hub\gamma_4^\prime-6\f{\dot{\hub}}{a\hub}\gamma_4+3aH_0\gamma_2^\prime+3H_0\gamma_2 \\
&&\tilde{W}_6^\prime=-4\l[\f{1}{2}\Omega^\prime+\gamma_5^\prime\r].
\ea
Let us notice that for models characterized by $\gamma_3\neq-\gamma_4$ and/or $\gamma_6\neq0$ the stability conditions need to be derived separately; while, the full set of perturbative equations evolved by EFTCAMB is still valid and can be used for a thorough investigation of linear perturbations, the derivation of the corresponding stability conditions  is work in progress. Therefore, for these cases the user has to be careful in setting appropriate stability conditions for the chosen model. Ho\v rava gravity belongs to this class of theories and in the current version of the code we implement specific stability conditions as explained in Section~\ref{Horava}.

Along with the physical stability conditions discussed above, and that are work in progress, we have a set of mathematical stability conditions which guarantee that the perturbations in the dark sector are stable. They are turn off by default, but can be turned on by the user providing a second layer of protection from unhealthy models. 
They are imposed via the $\pi-$equation and we call them mathematical (or classical) stability conditions. Rewriting the $\pi$ field equation as follows:
\begin{align}
& \left[A_1(\tau) + k^2 A_2(\tau)\right]\,\ddot{\pi} + \left[ B_1(\tau)+ k^2 B_2(\tau)\right]\,\dot{\pi} + C(\tau)\,\pi + k^2\,\left[D_1(\tau) + k^2D_2(\tau) \right]\,\pi + H_0 E(\tau,k) =0   \,,
\end{align}
the conditions that we impose read: 
\begin{itemize}
\item $A_1 + k^2 A_2 \neq 0$: well defined $\pi$ field equation;
\item No fast exponential growing of $\pi$ field perturbations:
\begin{itemize}
\item if $ B^2 -4(A_1+ k^2 A_2)( C +k^2D_1 + k^4D_2) > 0$ \\ 
then $ \left[ -B_1 - k^2 B_2 \pm \sqrt{B^2 -4(A_1+ k^2 A_2)( C +k^2D_1 + k^4D_2)} \right] / \left[ 2 (A_1 + k^2 A_2) \right] < H_0$
\item if $ B^2 -4(A_1+ k^2 A_2)( C +k^2D_1 + k^4D_2) < 0$ then $ - \left[ B_1 + k^2 B_2\right] / \left[ 2 (A_1 + k^2 A_2) \right] < H_0$  
\end{itemize}
\item $A_T \neq 0$: well defined tensor perturbations equation;
\end{itemize}

Let us notice that all the conditions based on scale dependent relations are enforced  from $k=0$ up to $k=k_{max}$, where $k_{max}$ is the maximum wavenumber that CAMB evolves.

The EFTCAMB code allows the user to have full control over the prior that are enforced by means of a set of flags contained in the parameter file \texttt{params$\_$EFT.ini} that will do the following:
\begin{itemize}
\item \texttt{EFT$\_$mathematical$\_$stability}: decides whether to enforce requirements of mathematical stability;
\item \texttt{EFT$\_$physical$\_$stability}: establishes whether to use physical viability conditions;
\item \texttt{EFT$\_$additional$\_$priors}: determines whether to use model specific priors on cosmological parameters;
\end{itemize}
%

\subsection{Initial Conditions}\label{SubSec:InitialConditions}
We assume that DE perturbations are sourced by matter perturbations at a sufficiently early time so that the theory is close to GR and initial conditions can be taken to be:
\begin{align}
\pi \l(\tau_\pi\r) =&\, -H_0 \f{E(\tau_\pi)}{C(\tau_\pi) +k^2 D(\tau_\pi)} \,,\nonumber \\
\dot{\pi} \l( \tau_\pi \r) =&\, H_0\l[\frac{E(\tau_\pi)}{(C(\tau_\pi)+k^2D(\tau_\pi))^2}(\dot{C}(\tau_\pi)+k^2\dot{D}(\tau_\pi))-\frac{\dot{E}(\tau_\pi)}{C(\tau_\pi)+k^2D(\tau_\pi)}\right] \,,
\end{align}
where $\tau_\pi$ is the time at which the code is switching from GR to DE/MG. This scheme is enforced here to speed up models that are too close to GR at early times.

Studying early DE/MG models requires the user to modify the parameter flag \verb|EFTturnonpiInitial| to a suitable value. Its default value is set to be  \verb|EFTturnonpiInitial|=0.01. The  initial conditions for matter components and curvature perturbations are set in the radiation dominated epoch ($a \sim 10^{-8}$). 

\section{Pure EFT models}\label{Sec:PureEFTmodels}
In the \textit{pure} EFT procedure once the background expansion history has been fixed, one has to specify the functional forms for the EFT functions.  EFTCAMB allows to choose among several models. We write them here just for $\Omega$ but the same time dependence can be assumed for any other EFT function $\gamma_1,\dots,\gamma_6$.
\begin{itemize}
\item[] Constant models: $\Omega(a)=\Omega_0$; 
\item[] Linear models: $\Omega(a)=\Omega_0 a$;
\item[] Power law models: $\Omega(a)=\Omega_0 a^s$;
\item[] Exponential models: $\Omega(a)=  \exp{\left(\Omega_0 a^s\right)}-1$.
\end{itemize}

The first option includes the minimal coupling, corresponding to $\Omega=0$; the linear model  can be thought of as a first order approximation of a Taylor expansion; while the power law is inspired by $f(R)$. There is also the possibility for the user to choose an arbitrary form of $\Omega/\gamma_i$  according to any ans$\ddot{\text{a}}$tz the user wants to investigate, defining them as a function of the scale factor and by specifying their derivatives with respect to the scale factor.  Of course the possibility to set all/some second order EFT functions to zero is included.

In the code we implemented a slot for user defined forms which can be easily spotted inside the file \verb|EFT_main.f90|.
Once the user defined form has been specified no other modifications to the code are required but numerical stability is not guaranteed.
Notice also that due to the structure of our modification it is possible to use, inside the definition of the EFT functions, cosmological parameters like $\Omega_\Lambda$ and $\Omega_m$. 

Code notation for pure EFT models:
\begin{itemize}
\item $\Omega_0 = \texttt{EFTOmega0}$, $s = \texttt{EFTOmegaExp}$;
\item $\gamma_1^0 = \texttt{EFTGamma10}$, $s = \texttt{EFTGamma1Exp}$;
\item $\gamma_2^0 = \texttt{EFTGamma20}$, $s = \texttt{EFTGamma2Exp}$;
\item $\gamma_3^0 = \texttt{EFTGamma30}$, $s = \texttt{EFTGamma3Exp}$;
\item $\gamma_4^0 = \texttt{EFTGamma40}$, $s = \texttt{EFTGamma4Exp}$;
\item $\gamma_5^0 = \texttt{EFTGamma50}$, $s = \texttt{EFTGamma5Exp}$;
\item $\gamma_6^0 = \texttt{EFTGamma60}$, $s = \texttt{EFTGamma6Exp}$.
\end{itemize}

\subsection{Horndeski implementation into EFTCAMB}\label{pureHorndeski}

The Horndeski gravity~\cite{Horndeski:1974wa} or  Generalized Galileons (GG)~\cite{Deffayet:2009mn} correspond to the most general scalar tensor theory with second order derivatives in the equations. As discussed in~\cite{Gleyzes:2013ooa,Bloomfield:2013efa} they are a subset of the models encompassed by the EFT framework which corresponds to the following conditions on the EFT functions (in terms of our convention):
\be\label{Horndeski_condition}
2\gamma_5=\gamma_3=-\gamma_4\,,\,\,\,\,\gamma_6=0
\ee
Models belonging to this class can be implemented into EFTCAMB following three different procedures as we discuss in what follows. 
\begin{itemize}
\item Pure EFT: if one is interested in investigating the general Horndeski class, rather than implementing a specific model within it, then one can opt to work directly with the four EFT functions that describe the Horndeski class in the EFT framework~\cite{Bloomfield:2013efa}, i.e. $\{\Omega(a), \gamma_1(a),\gamma_2(a),\gamma_3(a)\}$, after an expansion history has been chosen ($w_{\rm DE}$). In this case one has to parametrize the dependence of the EFT functions on time instead of specifying the functions  ($K,G_i$).  The limit of this procedure is that the user loses information about the corresponding theory as an inverse machinery to reconstruct the Horndeski functions $(K,G_i)[\phi,X]$ is not possible to implement.    
\item  EFT implementation of the parametrization proposed in~\cite{Bellini:2014fua}; the latter is a parametrization of Horndeski theory in terms  of five functions of time which are chosen to correspond to specific physical properties of the scalar d.o.f.. Let us notice that it is equivalent to the pure EFT approach, and hence it shares the same limit mentioned above, while perhaps being closer to the phenomenology and the observables in the specific choice of the five functions. See Section~\ref{RPH} for details. 
\item Full/designer Mapping approaches: one starts from a specific Horndeski/GG model, i.e. from a choice of the 4 unknown  functions $(K,G_i)[\phi,X]$ (where $i=3,4,5$ and $x\equiv -\partial_\mu \phi\partial^\mu\phi/2$) and in the case of designer mapping by specifying also the expansion history, i.e. $w_{\rm DE}$; proceeds solving the background field equations for this model and then, through the mapping described in~\cite{Gleyzes:2013ooa,Bloomfield:2013efa}, reconstructs the corresponding EFT functions. At this point EFTCAMB has all the ingredients to evolve linear perturbations. This approach emphasizes the choice of a specific theory and its implementation in EFTCAMB is work in progress.
\end{itemize}

Here we will focus on the pure EFT approach leaving the other options to the specific Sections. To use the code using the EFT approach restricted to Horndeski case, the user has to change the flag \verb|PureEFTHorndeski| setting it to true. Once this has been done, the code will  automatically restrict the number of the involved second order EFT functions according to eq.~(\ref{Horndeski_condition}). This condition will internally  fix the behaviour of the EFT functions $\gamma_4$, $\gamma_5$ and $\gamma_6$ so that the corresponding choices and parameters for these functions will be ignored. At this point the user can choose according to the model he/she wants to investigate, the behaviours of the three remaining EFT functions $\{\gamma_1,\gamma_2,\gamma_3\}$  and  $\{\Omega, w_{DE}\}$ to fix the background. Built-in models for these functions are already present in the code, see Section~\ref{Sec:PureEFTmodels}, as well as the possibility to implement the user own model.

\section{Alternative model-independent parametrizations in the EFT language} \label{Alternative}

Besides the EFT approach, in  literature there are several alternative model-independent parametrizations which allow to describe possible departures from General Relativity which come from a modification at the level of the action or directly by parametrizing the field equations by introducing new functions or parameters. Due to the versatility of the EFT approach and to the large number of DE/MG models that can be described within this framework, it is possible to completely cast most of these model-independent parametrizations in terms of EFT functions.   In this Section we will gradually implement the technical details of such alternative parametrizations. 

\subsection{RPH: ReParametrized Horndeski}\label{RPH}

As already discussed, there are three ways to implement the Horndeski/GG theory in EFTCAMB (See Section~\ref{pureHorndeski}). In the following we want to focus on the
implementation of the parametrization proposed in~\cite{Bellini:2014fua}, which is a built-in feature of the latest release (v2.0). Hereafter we will refer to this parametrization as ReParametrized Horndeski (RPH) as it is a model-independent parametrization of Horndeski theory in terms  of five functions of time defined in such a way that they  correspond to specific physical properties of the scalar d.o.f.. They correspond to  the expansion history (in terms of $w(a)$) and four functions of time $\{\tilde{M},\alpha_K,\alpha_B,\alpha_T\}$ . There is a fifth function, which is not independent but can  be derived from $\tilde{M}$ and  whose expression reads $\alpha_M= a \tilde{M}^\prime/(1+\tilde{M})$. As it will be clear in the following,  RPH is equivalent to the pure EFT approach.

As described in~\cite{Bellini:2014fua}  the correspondence between these functions and the EFT ones is (in our convention):
 \begin{eqnarray}
\f{M_*^2}{m_0^2}=1+ \tilde{M}&=& 1+\Omega+ \gamma_3 \, , \nn \\
\alpha_K&= & \f{\f{2ca^2}{m_0^2}+4H_0^2\gamma_1a^2}{(1+\Omega+ \gamma_3)\hub^2} \, ,\nn \\
\alpha_B&=&  +\f{1}{2}\f{a\gamma_2H_0+a\hub\Omega^\prime}{\hub(1+\Omega+ \gamma_3) } \,, \nn \\
\alpha_T&=& -\f{\gamma_3}{1+\Omega+ \gamma_3} \,.
\end{eqnarray}
Notice that we have redefined $\f{M_*^2}{m_0^2}=1+ \tilde{M}$ for numerical reasons. \\
To implement this parametrization into EFTCAMB we need first to invert the previous definitions in order to calculate $\{\Omega, \gamma_1\gamma_2,\gamma_3\}$ as functions of $\{\tilde{M},\alpha_K,\alpha_B,\alpha_T\}$:
\begin{align}
\Omega(a) &= (1+ \tilde{M})\alpha_T+ \tilde{M}\,, \nonumber \\
\gamma_1(a) &= \f{1}{4H_0^2a^2}\l[\alpha_K(1+ \tilde{M})\hub^2-\f{2ca^2}{m_0^2}\r]\,,  \nonumber \\
\gamma_2(a) &=  \f{1}{aH_0}\l[+2\alpha_B\hub (1+ \tilde{M})-a\hub\Omega^\prime\r] \,,  \nonumber \\
\gamma_3(a) &= -\alpha_T(1+ \tilde{M}) \,,  \nonumber \\
\gamma_4(a) &= -\gamma_3\,.
\end{align}
Then we can compute the following derived quantities that are needed for the equations in the code:
\begin{align}
\Omega'(a) &=  (1+\tilde{M})\alpha_T^\prime+ \tilde{M}^\prime(\alpha_T+1)\,,  \nonumber \\ 
\Omega''(a) &= 2\tilde{M}^\prime \alpha_T^\prime + (1+\tilde{M})\alpha_T^{\prime\prime}+\tilde{M}^{\prime\prime}(\alpha_T+1)\,,  \nonumber \\
\Omega'''(a) &=   3\tilde{M}^{\prime\prime}\alpha_T^\prime+3\tilde{M}^\prime\alpha_T^{\prime\prime}+ (1+\tilde{M})\alpha_T^{\prime\prime\prime}+\tilde{M}^{\prime\prime\prime}(\alpha_T+1)\,,  \nonumber \\
\gamma_1'(a) &= -\f{2}{a}\gamma_1+\f{1}{4H_0^2a^2}\l[\alpha_K^\prime(1+ \tilde{M})\hub^2+\alpha_K\tilde{M}^\prime\hub^2+2\alpha_K(1+ \tilde{M})\f{\dot{\hub}}{a}-\f{2c^\prime a^2}{m_0^2}-\f{4ca}{m_0^2}\r] \,,  \nonumber \\
\gamma_2'(a) &=  -\f{\gamma_2}{a} -\f{1}{aH_0}\l[-2(1+ \tilde{M})(\alpha_B^\prime\hub+\alpha_B \f{\dot{\hub}}{a\hub}) -2\alpha_B\hub \tilde{M}^\prime+ \hub\Omega^\prime+\f{\dot{\hub}}{\hub}\Omega^\prime+a\hub\Omega^{\prime\prime}\r] \,,  \nonumber \\
\gamma_3'(a) &= -(1+ \tilde{M})\alpha_T^\prime- \alpha_T \tilde{M}^\prime\,,  \nonumber \\
\gamma_4(a) &= -\gamma_3\,,  \nonumber \\
\gamma_4'(a) &= -\gamma_3^\prime\,,  \nonumber \\
\gamma_4''(a) &= -\gamma_3^{\prime\prime}= (1+ \tilde{M})\alpha_T^{\prime\prime}+2\alpha_T^\prime \tilde{M}^\prime + \alpha_T \tilde{M}^{\prime\prime} \,,  \nonumber \\
\gamma_5(a) &= \f{\gamma_3}{2} \,,  \nonumber \\
\gamma_5'(a) &=  \f{\gamma_3^\prime}{2}\,, \nonumber \\
\gamma_6(a) &= 0\,,  \nonumber \\
\gamma_6'(a) &=0 \,.
\end{align}

To use the  RPH  parametrization of Horndeski models, one has to set \verb|EFTflag=3| as described earlier (see Section~\ref{Sec:Code structure}) and then choose \verb|AltParEFTmodel=1|. Once this has been done, the user has to choose the behaviour for the four RPH functions of time $\{\tilde{M},\alpha_K,\alpha_B,\alpha_T\}$  by acting on the \verb|RPHmassPmodel|, \verb|RPHkineticitymodel|, \verb|RPHbraidingmodel| and \verb|RPHtensormodel| flags and for the expansion history by choosing $w_{DE}$. The built-in models allow to choose a constant (e.g.: $\alpha_K=\alpha_K^0$) and power law (e.g.: $\alpha_K = \alpha_K^0\, a^s$), or the  user can define by him/herself the behaviours for the respective functions according to the model he/she wants to investigate. The name of the parameters are then specified by:
\begin{itemize}
\item $\tilde{M}_0 = \verb|RPHmassP0|$, $s = \verb|RPHmassPexp|$;
\item $\alpha_{K}^0 = \verb|RPHkineticity0|$, $s = \verb|RPHkineticityexp|$;
\item $\alpha_{B}^0 = \verb|RPHbraiding0|$, $s = \verb|RPHbraidingexp|$;
\item $\alpha_{T}^0 = \verb|RPHtensor0|$, $s = \verb|RPHtensorexp|$;
\end{itemize}
By running parameter space explorations of these models we noticed that the viability priors~\cite{Bellini:2014fua} are particularly aggressive so we suggest to have some ideas of the parameter space of these models before studying them.

\section{Designer Mapping EFT models}\label{Sec:MatchingEFT}
The EFT framework allows to study a specific single field DE/MG model once the mapping into the EFT language is known. We refer the reader to~\cite{Gubitosi:2012hu,Bloomfield:2012ff,Gleyzes:2013ooa,Bloomfield:2013efa} for a complete list of the theories that can be cast in the EFT framework and  for an exhaustive theoretical treatment of models already mapped in this language. 

Once the user chooses the model of interest, a model-dependent flag solves the corresponding background equations for a given expansion history ($\Lambda$CDM, $w$CDM or CPL). Then, using the mapping into the EFT formalism, it reconstructs the corresponding EFT functions and, finally, has all the ingredients to evolve the full dynamical EFT perturbed equations.
Notice that in this case all the EFT functions are completely specified by the choice of the model and once the background equations are solved. 
In the current version,  EFTCAMB includes flags for $f(R)$ models and  minimally coupled quintessence. The background equations are solved through the use of the designer approach specific to $f(R)$~\cite{Song:2006ej,Pogosian:2007sw} (see the following Subsection for implementation details). In the future, new flags implementing the background equations and the mapping for other DE/MG models of interest that are included in the EFT formalism will be added. A detailed diagram of the \textit{mapping} EFT case is shown in Figure~\ref{Fig:StructureFlowChart}.
\subsection{Designer $f(R)$}\label{SubSec:DesignerFR}
We consider the following action in Jordan frame
\begin{equation} 
S=\int d^4x \sqrt{-g} \l(R+f(R)\r)+S_m, 
\end{equation}
where $f$ is a generic function of the Ricci scalar, $R$, and $S_m$ indicates the action for matter fields which, in this frame, are minimally coupled to gravity.
$f(R)$ models can  be mapped into the EFT language via the following relations~\cite{Gubitosi:2012hu}:
\begin{align} \label{Eq:MatchingFR}
\Lambda(t)=\frac{m_0^2}{2}\l[f- Rf_R \r] \hspace{0.5cm};\hspace{0.5cm} c(t)=0 \hspace{0.5cm};\hspace{0.5cm} \Omega(t)=f_R,
\end{align}
where $f_R\equiv\frac{df}{dR}$. 

It is well known that  given the higher order nature of $f(R)$ gravity, there is enough freedom to reproduce any desired expansion history. When dealing with perturbations, it is common to adopt the \emph{designer} approach which consists in fixing the expansion history and solving the Friedman equation as a second order differential equation in $\ln a$ for the function $f[R(a)]$~\cite{Song:2006ej,Pogosian:2007sw}. This procedure yields, for each chosen background, a family of $f(R)$ models that can be labelled by the value of $f_R\equiv df/dR$ today, or analogously, by the present value of the mass scale of the scalaron
\be\label{B}
B\equiv\f{f_{RR}}{1+f_R}\f{\hub\dot{R}}{\dot{\hub}-\hub^2},
\ee
One further needs to impose certain viability conditions on the resulting models in order to have stable and viable cosmologies~\cite{Pogosian:2007sw}.
We refer the reader to~\cite{Song:2006ej,Pogosian:2007sw} for the details of the designer approach to $f(R$) models. In the following Subsection we shall present some of the technical aspects of its implementation in EFTCAMB.

Following~\cite{Pogosian:2007sw}, let us define the dimensionless quantities:
\begin{align}
y\equiv \frac{f(R)}{H_0^2}, \hspace{0.8cm} E \equiv \frac{H^2}{H_0^2}, \hspace{0.8cm} \frac{R}{H_0^2} \equiv 3\left(4 E +E' \right),\hspace{0.8cm}E_{m,r}\equiv\f{\rho_{m,r}}{\rho_c^0}\,,
\end{align}
where $\rho_c^0\equiv 3H_0^2M_P^2$ is the critical density today and, \emph{in this Section only, primes indicate derivation with respect to $\ln a$} ( to  be not confused with the primes indicating derivatives with respect to $a$ in the previous Sections).
 
Furthermore, let us introduce an effective energy density, $E_{\rm{eff}}$, and an effective equation of state, $w_{\rm eff}$, which allow one to set the desired expansion history:
\begin{align}\label{E_Hubble}
E&\equiv E_m+E_r+E_{\rm{eff}}\nonumber\\
E_{\rm{eff}}&= \Omega_{DE}\exp\left[-3x+3\int_a^1 w_{\rm{eff}}(\tilde{a}) d\ln\tilde{a} \right]\,.
\end{align} 
In terms of these dimensionless quantities the scalaron mass scale reads:
\begin{align}\label{B_E}
B =\frac{2}{3(1+f_R)}\frac{1}{4 E'+E''}\frac{E}{E'}\left(y'' -y'\frac{4 E''+E'''}{4 E' + E''} \right)\,.
\end{align}
The designer approach will then consist in choosing one of the expansion histories in Section~\ref{SubSec:DeEOS} and solving the following equation for $y(x)$
\be\label{designer_fR_EFTCAMB}
y''-\l(1+\f{E'}{2E}+\f{R''}{R'}\r)y'+\f{R'}{6H_0^2E}y=-\f{R'}{H_0^2E}E_{\rm eff}\,,
\ee
with appropriate boundary conditions that allow us to select the growing mode, as described in~\cite{Pogosian:2007sw}. 
The outcome will be a family of models labelled by the present day value, $B_0$, of~(\ref{B_E}).  \\
In order to implement eq.~(\ref{designer_fR_EFTCAMB}) in the code we need to define the following quantities:
\begin{align}
E_{\rm eff} \equiv \Omega_{\Lambda} e^{-3g(x)} \, \Rightarrow \, g(x) &= \int_1^{e^x} \frac{1+w_{\rm DE}(\tilde{a})}{\tilde{a}} d\tilde{a} \,, \nonumber \\
g'(x) &= 1+w_{\rm DE} \,, \nonumber \\
g''(x) &= e^x \frac{d w_{\rm DE}}{da} \,, \nonumber \\
g^{(3)}(x) &= e^x \frac{d w_{\rm DE}}{da} + e^{2x} \frac{d^2 w_{\rm DE}}{da^2} \,,
\end{align}
where we have introduced the variable $x\equiv \ln a$.
These definitions allow us to rewrite $E(x)$ and its derivatives as follows:
\begin{align}
E' =& -3\Omega_m e^{-3x}-4\Omega_re^{-4x} -3\Omega_\Lambda e^{-3g(x)}g'(x) \, ,\nonumber\\
E''=& 9 \Omega_m e^{-3x} + 16\Omega_r e^{-4x} -3\Omega_\Lambda e^{-3g(x)}\left(g''(x) -3g'(x)^2 \right) \, ,\nonumber\\
E''' =& -27 \Omega_m e^{-3x} -64 \Omega_r e^{-4x} -3\Omega_\Lambda e^{-3g(x)}\left(g^{(3)}(x) -9 g'(x)g''(x) +9 g'(x)^3\right)\,.
\end{align}
Let us notice that massive neutrinos are consistently implemented in the designer approach. For a detailed treatment of their inclusion in the above equations we refer the user to~\cite{Hu:2014sea} and references therein for an historical background.
Once we have solved the background according to the above procedure, there remains only to map the solution into the EFT language. To this extent we have:
\begin{align}
\Omega(a)\equiv f_R(a)=\f{y'}{3(4E' + E'')}\,.
\end{align}
For the code purposes we need also the derivatives of this function with respect to $a$, and it turns useful to input their analytical expressions directly in the code, rather than having the code evaluate them numerically. We have:

\begin{align}\label{Omega_eqs}
\frac{d\Omega}{da} =& \frac{e^{-x} \left(-E'' (6 E_{\rm eff}+y)+E'\left(y'-4 (6 E_{\rm eff}+y)\right)+2 E y'\right)}{6 E \left(E''+4 E'\right)}\, , \nonumber\\
\frac{d^2\Omega}{da^2} =& \frac{e^{-2x} \left(E'\left(E'\left(24 E_{\rm eff} -y' +4 y \right)-6 E \left(8 E_{\rm eff}' +y' \right)\right)+E''  \left(E'  (6 E_{\rm eff} +y )-12 E  E_{\rm eff}' \right)\right)}{12 E ^2 \left(E'' +4 E' \right)}\, ,\nonumber \\
\frac{d^3\Omega}{da^3} =& \frac{e^{-3 x}}{24 E^3 \left(E'' +4 E' \right)} \bigg[2 E  \left(E^{\prime\prime\, 2} (6 E_{\rm eff} +y )+4 E^{\prime \, 2} \left(18 E_{\rm eff}' +6 E_{\rm eff} +2 y' +y \right)+E'  E''  \left(18 E_{\rm eff}' +30 E_{\rm eff} -y' +5 y \right)\right)\, , \nonumber\\
& +12 E^2 \left(E''  \left(-2 E_{\rm eff}'' +4 E_{\rm eff}' -y' \right)+E'  \left(-8 E_{\rm eff}'' +16 E_{\rm eff}' +y' \right)\right)+3 E^{\prime\,2} \left(E'  \left(y' -4 (6 E_{\rm eff} +y )\right)-E''  (6 E_{\rm eff} +y )\right)\bigg]\, .
\end{align}

At this point, having $H(a)$ and $\Omega(a)$ at hand, one can go back to the general treatment of the background in the EFT formalism (setting $w_{\rm DE}=w_{\rm eff}$), and use the designer EFT described in Section~\ref{SubSec:EFTDesigner} to determine $c$ and $\Lambda$. However, for a matter of numerical accuracy, it is better to determine these functions via the mapping too. We have:
\begin{align}
\frac{c a^2}{m_0^2} = &0,\hspace{0.8cm}  \frac{\dot{c}a^2}{m_0^2} = 0 \, , \nonumber\\
\frac{\Lambda a^2}{m_0^2} =& a^2 \frac{\mathcal{H}_0^2}{2}\left[y - 3 f_R\left(4 E + E' \right) \right]\, ,\hspace{0.8cm} \frac{\dot{\Lambda} a^2}{m_0^2} = -\frac{3}{2}\mathcal{H}_0^2\, \mathcal{H}\left[a^3\frac{d \Omega}{da}\left( 4E + E'\right) \right]\, .
\end{align}

The user can select the designer-f(R) model by setting  \verb|mappingEFTmodel=1| in the  designer flag (\verb|EFTflag=2|)  as described earlier (see Section~\ref{Sec:Code structure}).

\subsection{Designer minimally coupled quintessence models}\label{SubSec:DesignerMc5e}
Minimally coupled quintessence models correspond to setting all the EFT functions $\Omega$ and $\gamma$'s to zero, while using an effective dark energy equation of state different from $-1$. \\
These models are specified by the action:
\begin{align}
S = \int{}d^4x\sqrt{-g}\l[\frac{m_0^2}{2}R-\frac{(\nabla\phi)^2}{2}-V(\phi)\r] \,,
\end{align}
and can be mapped into the EFT framework by~\cite{Gubitosi:2012hu, Bloomfield:2012ff}:
\begin{align}
c(\tau)  =&\,\, \frac{1}{2 a^2} \dot{\phi}_0^2 \,\,, \nonumber \\
\Lambda(\tau) =&\,\, \frac{1}{2a^2} \dot{\phi}_0^2 -V\left(\phi_0(\tau)\right) \,\,,
\end{align}
where $\phi_0$ is the background value of the quintessence field and $V\left(\phi_0\right)$ is the quintessence potential. \\
It can be shown that fixing the background expansion history by a designer approach results in:
\begin{align}
\frac{c a^2}{m_0^2} =&\,\, \frac{1}{2}\frac{a^2 \rho_{\rm DE}}{m_0^2}(1+w_{\rm DE})\, , & &   [\text{Mpc}^{-2}] \nonumber \\
 \frac{\Lambda a^2}{m_0^2} =&\,\,  w_{\rm DE}\frac{a^2\rho_{\rm DE}}{m_0^2}\, , & & [\text{Mpc}^{-2}]  
\end{align}
and their time derivatives can be read from equation~(\ref{Eq:DesignerCLambda}). 

The user can select the minimally coupled quintessence models by setting  \verb|mappingEFTmodel=2| in the  designer flag (\verb|EFTflag=2|) as described earlier (see Section~\ref{Sec:Code structure}). Once this has been done, the user has to choose  only the behaviour of $w_{\rm DE}$.

\section{Full EFT Mapping}\label{Fullmapping}

As already discussed, the EFT framework offers a unifying language for all single field models of DE and MG; once a given model is mapped into the EFT language, i.e. the corresponding EFT functions are determined, one does not need to derive  lengthy  perturbation equations specific for that model. Rather, the general perturbation equations for the EFT action, which are implemented in EFTCAMB, can be used. All the necessary input are the background evolution and the EFT functions as functions of the scale factor. There are two ways in which the latter can be determined: via the designer approach discussed in the previous Section, or solving for the background evolution of the given model, determining the Hubble parameter and  the time evolution of the EFT functions. Once these are worked out, they are passed to the main code  which solves  the full perturbative equations.

The users can choose this branch by setting \verb|EFTflag=4|. The current built-in model is low-energy Ho\v rava gravity, implemented as described in~\cite{Frusciante:2015maa}, which can be selected by choosing \verb|FullMappingEFTmodel=1|.

\subsection{Low-energy Ho\v rava gravity}\label{Horava}

A thorough discussion of the theory considered and  analysis of its cosmological implications via EFTCAMB can be found in~\cite{Frusciante:2015maa}. Here we will report the action and few important details about its implementation in the code. The action that we consider corresponds to  the low-energy Ho\v rava gravity~\cite{Blas:2009qj}
\be\label{Horavaaction}
\mathcal{S}=\f{m_0^2}{(2\xi-\eta)}\int{}d^4x\sqrt{-g}\left[K_{ij}K^{ij}-\lambda K^2 -2 \xi\bar{\Lambda}+\xi \mathcal{R}+\eta a_i a^i\right]
\ee
where $K_{ij}$ and $K$ are the extrinsic curvature and its trace and $\mathcal{R}$ is the  Ricci scalar of the three-dimensional space-like hypersurfaces. The coefficients $\lambda$, $\eta$, $\xi$  are running coupling constants, while $\bar{\Lambda}$ is the ''bare'' cosmological constant,  $a_i \equiv \partial_i \mbox{ln} N$ being $N$ the lapse function in the ADM formalism. 
Let us now introduce the following definitions useful for numerical computations
\begin{align}
\tilde{\xi} = \xi - 1  \hspace{0.5cm};\hspace{0.5cm} \tilde{\lambda} = \lambda -1   
\label{redef}
\end{align}

Action (\ref{Horavaaction}) can be mapped in EFT formalism as follows~\cite{Frusciante:2015maa}
\begin{align}
& \Omega = \f{\eta}{2+2\tilde{\xi}-\eta} \,\,;\,\, \Omega' = \Omega'' = \Omega''' = 0, \nonumber \\
& \gamma_1 = \frac{1}{2 H_0^2a^2 }\f{ 2\tilde{\xi}-3\tilde{\lambda}}{(2+2\tilde{\xi}-\eta)}\left( \dot{\hub} -\hub^2 \right) \,\,, \nonumber\\
& \gamma_1' = \frac{1}{2 H_0^2a^3 }\f{ 2\tilde{\xi}-3\tilde{\lambda}}{(2+2\tilde{\xi}-\eta)}\left( \frac{\ddot{\hub}}{\hub} -4\dot{\hub} +2\hub^2 \right),\nonumber \\
& \gamma_2 = 0 \,\,;\,\, \gamma_2' = \gamma_2'' = 0, \nonumber \\
& \gamma_3 = 2\f{\tilde{\lambda} -\tilde{\xi}}{(2+2\tilde{\xi}-\eta)}  \,\,;\,\, \gamma_3' = \gamma_3'' = 0, \nonumber \\
& \gamma_4 = \f{2\tilde{\xi}}{(2+2\tilde{\xi}-\eta)} \,\,;\,\, \gamma_4' = \gamma_4'' = 0, \nonumber \\
& \gamma_5 = 0 \,\,;\,\, \gamma_5' = \gamma_5'' = 0, \nonumber \\
& \gamma_6 = \frac{\eta}{4(2+2\tilde{\xi}-\eta)} \,\,;\,\, \gamma_6' = \gamma_6'' = 0, \nonumber \\
& \frac{ca^2}{m_0^2} = -\f{ 2\tilde{\xi}-3\tilde{\lambda}}{(2+2\tilde{\xi}-\eta)}\left( \dot{\hub} -\hub^2 \right), \nonumber \\
& \frac{\dot{c}a^2}{m_0^2} = -\f{ 2\tilde{\xi}-3\tilde{\lambda}}{(2+2\tilde{\xi}-\eta)}\left( \ddot{\hub} -4\hub\dot{\hub} +2\hub^3 \right), \nonumber \\
& \frac{\Lambda a^2}{m_0^2} = -3\l(\Omega_{DE}^0-1+ \f{3\tilde{\lambda}+2}{2\tilde{\xi}+2-\eta}\r)H_0^2 a^2 -2\f{ 2\tilde{\xi}-3\tilde{\lambda}}{(2+2\tilde{\xi}-\eta)}\left( \frac{\hub^2}{2} +\dot{\hub} \right), \nonumber \\
& \frac{\dot{\Lambda} a^2}{m_0^2} = -2\f{ 2\tilde{\xi}-3\tilde{\lambda}}{(2+2\tilde{\xi}-\eta)} \left( \ddot{\hub} -\hub\dot{\hub} -\hub^3 \right).
\end{align}

The background of Ho\v rava gravity is fully solved without imposing \textit{a priori} any condition. In detail, we implemented in EFTCAMB and solved the following equation which describes the  background  evolution:
\be
\hub^2=\f{(2\tilde{\xi}+2-\eta)}{3\tilde{\lambda}+2}a^2\l\{\f{8\pi G_N}{3}(\rho_m+\rho_{\nu})+\l[\Omega_{DE}^0-1+ \f{3\tilde{\lambda}+2}{2\tilde{\xi}+2-\eta}\r]H_0^2 \r\}
\ee
where we have used the following relation
\be
\bar{\Lambda}= 3\f{2\xi-\eta}{2\xi}\l[\Omega_{DE}^0-1+\f{3\lambda-1}{2\xi-\eta}\r]H_0^2,
\ee
to eliminate the $\bar{\Lambda}$ parameter from the evolution equation in favour of $\Omega_{DE}^0$. As usual the density of the massive neutrinos is implemented as explained in~\cite{Hu:2014sea}. Moreover, we also computed the  derivatives of the Hubble function:
\ba
&&\hub^2=\f{(2\tilde{\xi}+2-\eta)}{3\tilde{\lambda}+2}a^2\l\{\f{8\pi G_N}{3}(\rho_m+\rho_{\nu})+\l[\Omega_{DE}^0-1+ \f{3\tilde{\lambda}+2}{2\tilde{\xi}+2-\eta}\r]H_0^2 \r\} \nn \\
&&\dot{\hub}= -\f{(2\tilde{\xi}+2-\eta)}{3\tilde{\lambda}+2}a^2\l[-\l(\Omega_{DE}^0-1+ \f{3\tilde{\lambda}+2}{2\tilde{\xi}+2-\eta}\r)H_0^2+\f{4\pi G_N}{3}(3P_{\nu}+\rho_{\nu})+\f{4\pi G_N}{3}(1+3w_m)\rho_m\r]  \nn \\
&&\ddot{\hub}=-\f{(2\tilde{\xi}+2-\eta)}{3\tilde{\lambda}+2}a^2\l[-2\l(\Omega_{DE}^0-1+ \f{3\tilde{\lambda}+2}{2\tilde{\xi}+2-\eta}\r)H_0^2\hub+\f{8\pi G_N}{3}(\f{3}{2}P_{\nu}\hub-\f{1}{2}\rho_{\nu}\hub+\f{3}{2}\dot{P}_{\nu}) \r. \nn \\&&\l.-\f{4\pi G_N \hub}{3}(1+6w_m +9w_m^2)\rho_m\r].
\ea
Ho\v rava gravity belongs to the class of theory for which EFTCAMB has not yet the appropriate general stability conditions, therefore for this model we implemented the stability conditions found in~\cite{Frusciante:2015maa} which, in code notation, read:
\be
\tilde{\lambda}>0\, \qquad 0<\eta<2\tilde{\xi}+2. 
\ee
These conditions are imposed by default and they become viability priors when using EFTCosmoMC.

To investigate  the low-energy  Ho\v rava gravity, the user has to set \verb|EFTflag=4| as described earlier (see Section~\ref{Sec:Code structure}) and then choose \verb|FullmappingEFTmodel=1|. At this point the user can study the model for which all the three parameters appearing in the action can vary. If one is interested in investigating the case for which the theory evades the Solar System constraints the user has to set \verb|HoravaSolarSystem=2|. For details about the physics of the two models see~\cite{Frusciante:2015maa}.

\newpage 

\section*{Change Log}\label{Sec:ChangeLog}

\begin{itemize}
\item Version 3.0 (Sep17):
\begin{itemize}
\item General restructuring of the notes to reflect new code structure;
\item Several minor typo fixed;
\end{itemize}
\item Version 2.0 (Oct15):
\begin{itemize}
\item EFTCAMB/EFTCosmoMC compatible with the Planck likelihood--PLC2.0;
\item Change in the parametrization of the $\alpha$s;
\item Added a section with the implementation of the Horndeski parametrization by using the pure EFT approach: Pure Horndeski;
\item Added a section with the implementation of alternative model-independent parametrizations in terms of EFT functions. Built-in: ReParametrized Horndeski (RPH);
\item Added mathematical stability requirements on the equation for the $\pi$-field; 
\item Added physical stability requirements  for the class of theories belonging to GLPV;
\item Added a section with the implementation of a full EFT mapping model: low energy Ho\v rava gravity;
\item Typos fixed in the $\pi$-field equation coefficients, which involve the $\gamma_4, \gamma_6$ functions.
\end{itemize}
\item Version 1.1 (Oct14):
\begin{itemize}
\item Added a section for DE equation of state parametrizations;
\item Added a section on tensor modes;
\item Added a section on viability priors;
\item Added a section on designer minimally coupled quintessence models;
\item Linear tensor Perturbations typo fixed, which involve ONLY second order operator;
\item EFTCAMB sources: LSS number counts, LSS weak lensing and all CMB cross-correlation functions in DE/MG models.
\end{itemize}
\item Version 1.0 (May14): first version of EFTCAMB/EFTCosmoMC.
\end{itemize}
%
\acknowledgments
We are grateful to Carlo Baccigalupi, Nicola Bartolo, Jolyon Bloomfield, Matteo Calabrese, Paolo Creminelli, Antonio De Felice, J\'er\^{o}me Gleyzes, Alireza Hojjati, Martin Kunz, Stefano Liberati, Matteo Martinelli, Sabino Matarrese, Ali Narimani, Simone Peirone, Valeria Pettorino, Federico Piazza, Levon Pogosian, Filippo Vernizzi, Bo Yu and Gong-Bo Zhao for useful conversations. We are indebted to Luca Heltai and Riccardo Valdarnini for help with numerical algorithms and to Jorgos Papadomanolakis and Daniele Vernieri for help in developing some theoretical aspects.

The research of BH is partially supported by the Chinese National Youth Thousand Talents Program, the Fundamental Research Funds for the Central Universities under the reference No. 310421107 and Beijing Normal University Grant under the reference No. 312232102. The work related to previous versions of these notes was supported by: the Dutch Foundation for Fundamental Research on Matter (FOM).
The research of MR is supported by U.S. Dept. of Energy contract DE-FG02-13ER41958. The work related to previous versions of these notes was supported by: the SISSA PhD Fellowship and the INFN-INDARK initiative (09/2012-08/2014).
The research of NF is currently supported by  Funda\c{c}\~{a}o para a  Ci\^{e}ncia e a Tecnologia (FCT) through national funds  (UID/FIS/04434/2013) and by FEDER through COMPETE2020  (POCI-01-0145-FEDER-007672). The  work related to previous versions of these notes was supported by: the European Research Council under the European Community's Seventh Framework Programme (FP7/2007-2013, Grant Agreement No.~307934), the European Research Council under the European Union’s Seventh Framework Programme (FP7/2007-2013) / ERC Grant Agreement n. 306425 “Challenging General Relativity”, the SISSA PhD Fellowship and INFN. 
The research of AS is currently supported by the Netherlands Organization for Scientific Research (NWO/OCW) and the D-ITP consortium, a program of the Netherlands Organisation for Scientific Research (NWO) that is funded by the Dutch Ministry of Education, Culture and Science (OCW). The  work related to previous versions of these notes was supported by the SISSA Excellence Grant and the INFN-INDARK  initiative (09/2012-08/2014).
NF and AS acknowledge the COST Action  (CANTATA/CA15117), supported by COST (European Cooperation in  Science and Technology).

%



\begin{thebibliography}{999}


\bibitem{Gubitosi:2012hu} 
  G.~Gubitosi, F.~Piazza and F.~Vernizzi,
JCAP {\bf 1302}, 032 (2013),
   [arXiv:1210.0201 [hep-th]]. 
 
\bibitem{Bloomfield:2012ff} 
  J.~K.~Bloomfield, \'E. \'E. ~Flanagan, M.~Park and S.~Watson,
JCAP {\bf 1308}, 010  (2013),
  [arXiv:1211.7054 [astro-ph.CO]].  
	
\bibitem{Gleyzes:2013ooa} 
  J.~Gleyzes, D.~Langlois, F.~Piazza and F.~Vernizzi,
  JCAP {\bf 1308}, 025 (2013),
   [arXiv:1304.4840 [hep-th]].

\bibitem{Bloomfield:2013efa} 
  J.~Bloomfield,
  JCAP {\bf 1312}, 044 (2013)
  [arXiv:1304.6712 [astro-ph.CO]].
  
\bibitem{Hu:2013twa} 
  B.~Hu, M.~Raveri, N.~Frusciante and A.~Silvestri,
  Phys.\ Rev.\ D {\bf 89}, no. 10, 103530 (2014)
  [arXiv:1312.5742 [astro-ph.CO]].

\bibitem{Raveri:2014cka} 
  M.~Raveri, B.~Hu, N.~Frusciante and A.~Silvestri,
  Phys.\ Rev.\ D {\bf 90}, no. 4, 043513 (2014)
  [arXiv:1405.1022 [astro-ph.CO]].
\bibitem{CAMB}
http://camb.info \,.

\bibitem{Lewis:1999bs} 
  A.~Lewis, A.~Challinor and A.~Lasenby,
  Astrophys.\ J.\  {\bf 538}, 473 (2000),
  [astro-ph/9911177].
	
\bibitem{Lewis:2002ah}
  A.~Lewis and S.~Bridle,
  Phys.\ Rev.\ D {\bf 66}, 103511 (2002)
  [astro-ph/0205436].
	
\bibitem{Zumalacarregui:2016pph} 
  M.~Zumalacárregui, E.~Bellini, I.~Sawicki, J.~Lesgourgues and P.~G.~Ferreira,
  JCAP {\bf 1708}, no. 08, 019 (2017)
  doi:10.1088/1475-7516/2017/08/019
  [arXiv:1605.06102 [astro-ph.CO]].
 
\bibitem{Huang:2015srv} 
  Z.~Huang,
  Phys.\ Rev.\ D {\bf 93}, no. 4, 043538 (2016)
  doi:10.1103/PhysRevD.93.043538
  [arXiv:1511.02808 [astro-ph.CO]].
  
\bibitem{Bellini:2017avd} 
  E.~Bellini {\it et al.},
  arXiv:1709.09135 [astro-ph.CO].
 
\bibitem{Blas:2012vn} 
  D.~Blas, M.~M.~Ivanov and S.~Sibiryakov,
  JCAP {\bf 1210}, 057 (2012)
  doi:10.1088/1475-7516/2012/10/057
  [arXiv:1209.0464 [astro-ph.CO]].
  
\bibitem{Horndeski:1974wa} 
  G.~W.~Horndeski,
  Int.\ J.\ Theor.\ Phys.\  {\bf 10}, 363 (1974).
	
\bibitem{Ma:1995ey} 
  C.~-P.~Ma and E.~Bertschinger,
  Astrophys.\ J.\  {\bf 455}, 7 (1995)
  [astro-ph/9506072].
	
	
	\bibitem{cambnotes}
http://cosmologist.info/notes/CAMB.pdf

\bibitem{Hu:2014sea} 
  B.~Hu, M.~Raveri, A.~Silvestri and N.~Frusciante,
  Phys.\ Rev.\ D {\bf 91}, no. 6, 063524 (2015)
  [arXiv:1410.5807 [astro-ph.CO]].


	
\bibitem{Chevallier:2000qy} 
  M.~Chevallier and D.~Polarski,
  Int.\ J.\ Mod.\ Phys.\ D {\bf 10}, 213 (2001),
 [gr-qc/0009008].
	
\bibitem{Linder:2002et} 
  E.~V.~Linder,
  Phys.\ Rev.\ Lett.\  {\bf 90}, 091301 (2003),
  [astro-ph/0208512].

\bibitem{Jassal:2004ej} 
  H.~K.~Jassal, J.~S.~Bagla and T.~Padmanabhan,
  Mon.\ Not.\ Roy.\ Astron.\ Soc.\  {\bf 356}, L11 (2005)
  [astro-ph/0404378].
  
\bibitem{Jassal:2006gf} 
  H.~K.~Jassal, J.~S.~Bagla and T.~Padmanabhan,
  Mon.\ Not.\ Roy.\ Astron.\ Soc.\  {\bf 405}, 2639 (2010)
  [astro-ph/0601389].

\bibitem{Hu:2014ega} 
  Y.~Hu, M.~Li, X.~-D.~Li and Z.~Zhang,
  Sci.\ China Phys.\ Mech.\ Astron.\  {\bf 57}, 1607 (2014)
  [arXiv:1401.5615 [astro-ph.CO]].
	
	

  
\bibitem{Amendola:2014wma} 
  L.~Amendola, G.~Ballesteros and V.~Pettorino,
  Phys.\ Rev.\ D {\bf 90}, no. 4, 043009 (2014)
  [arXiv:1405.7004 [astro-ph.CO]].


\bibitem{Raveri:2014eea} 
  M.~Raveri, C.~Baccigalupi, A.~Silvestri and S.~Y.~Zhou,
  Phys.\ Rev.\ D {\bf 91}, no. 6, 061501 (2015)
  [arXiv:1405.7974 [astro-ph.CO]].
  
\bibitem{Song:2006ej} 
  Y.~-S.~Song, W.~Hu and I.~Sawicki,
  Phys.\ Rev.\ D {\bf 75}, 044004 (2007),
  [astro-ph/0610532].
  
\bibitem{Pogosian:2007sw} 
  L.~Pogosian and A.~Silvestri,
  Phys.\ Rev.\ D {\bf 77}, 023503 (2008),
  [Erratum-ibid.\ D {\bf 81}, 049901 (2010)],\,\,\,\,\,\,\,\,\,\,\,\,\,
  [arXiv:0709.0296[astro-ph]].
	


\bibitem{Deffayet:2009mn} 
  C.~Deffayet, S.~Deser and G.~Esposito-Farese,
  Phys.\ Rev.\ D {\bf 80}, 064015 (2009)
  [arXiv:0906.1967 [gr-qc]].

\bibitem{Bellini:2014fua} 
  E.~Bellini and I.~Sawicki,
  JCAP {\bf 1407}, 050 (2014)
  [arXiv:1404.3713 [astro-ph.CO]].
	
\bibitem{Gleyzes:2014dya} 
  J.~Gleyzes, D.~Langlois, F.~Piazza and F.~Vernizzi,
  Phys.\ Rev.\ Lett.\  {\bf 114}, no. 21, 211101 (2015)
  [arXiv:1404.6495 [hep-th]].
	
\bibitem{Kase:2014cwa} 
  R.~Kase and S.~Tsujikawa,
  Int.\ J.\ Mod.\ Phys.\ D {\bf 23}, no. 13, 1443008 (2015)
  [arXiv:1409.1984 [hep-th]].
	
\bibitem{Frusciante:2015maa} 
  N.~Frusciante, M.~Raveri, D.~Vernieri, B.~Hu and A.~Silvestri,
  arXiv:1508.01787 [astro-ph.CO].

\bibitem{Blas:2009qj} 
  D.~Blas, O.~Pujolas and S.~Sibiryakov,
  Phys.\ Rev.\ Lett.\  {\bf 104}, 181302 (2010)
  [arXiv:0909.3525 [hep-th]].

  \end{thebibliography}
\end{document}